\documentclass[twocolumn, twocolappendix, numberedappendix]{openjournal}
\usepackage{natbib}
\usepackage{graphicx,amsmath,amssymb,amstext}
\usepackage{amsbsy,amsfonts,amsthm,xcolor}
\usepackage{aas_macros}
\usepackage[colorlinks,linkcolor=blue,citecolor=blue,urlcolor=blue ]{hyperref}
\usepackage[utf8]{inputenc}
\usepackage{float}
\usepackage[caption=false]{subfig}
\usepackage{upgreek}
\usepackage{enumitem}

\newcommand{\eagle}{\mbox{EAGLE}}

\newcommand{\jwst}{\mbox{\it JWST}}
\newcommand{\flares}{\mbox{FLARES}}

\newcommand{\synthesizer}{\mbox{\textsc{synthesizer}}}
\newcommand{\cloudy}{\mbox{\texttt{Cloudy}}}

\newcommand{\bptnii}{\mbox{BPT-[N\sc{ii}]}}
\newcommand{\oii}{\mbox{[O\sc{ii}]}}
\newcommand{\oiii}{\mbox{[O\sc{iii}]}}
\newcommand{\hii}{\mbox{H\sc{ii}}}
\newcommand{\ha}{\mbox{H$\alpha$}}
\newcommand{\hb}{\mbox{H$\beta$}}

\newcommand{\Msun}{{\rm M_{\odot}}}

\newcommand{\eg}{$\textnormal{e.g.,}$}
\newcommand{\ie}{$\textnormal{i.e.}$}

\defcitealias{lovell2020}{\flares\,I}

\begin{document}

\title{Stellar photoionisation modelling in SYNTHESIZER}
\author{Aswin P. Vijayan$^{1\star}$}\email{$^{\star}$email: aswinpvijayan@gmail.com}
\author{Stephen M. Wilkins$^{1\dagger}$}\email{$^{\dagger}$email: S.Wilkins@sussex.ac.uk}
\author{Sophie L. Newman$^{2}$}
\author{Christopher C. Lovell$^{3,4}$}
\author{William J. Roper$^{1}$}
\author{Sabrina Berger$^{5}$}
\author{Thomas Harvey$^{6}$}
\author{Jack C. Turner$^{1}$}

\affiliation{$^1$Astronomy Centre, University of Sussex, Falmer, Brighton BN1 9QH, UK}
\affiliation{$^{2}$Institute of Cosmology \& Gravitation, University of Portsmouth, Portsmouth, PO1 3FX, UK}
\affiliation{$^{3}$Kavli Institute for Cosmology, Madingley Road, Cambridge CB3 0HA, UK}
\affiliation{$^{4}$Institute of Astronomy, Madingley Road, Cambridge CB3 0HA, UK}
\affiliation{$^{5}$School of Physics, University of Melbourne, Parkville, VIC 3010, Australia}
\affiliation{$^{6}$Jodrell Bank Centre for Astrophysics, University of Manchester, Oxford Road, Manchester M13 9PL, UK}

\begin{abstract}

Emission from photoionised gas surrounding young stellar populations (\hii\ regions) provides critical diagnostics of the physical conditions in star forming galaxies. This emission constrains the gas properties, the nature of ionising sources, and generates essential features for determining galaxy redshifts.
To leverage spectroscopic observations to test galaxy formation models, it is essential to incorporate these emissions into synthetic datasets. 
Here, we present the integration of photoionised gas emission into the \textsc{synthesizer} package (\href{https://synthesizer-project.github.io/}{synthesizer-project.github.io}) and demonstrate its application. 
We quantify the impact of key modelling assumptions—including stellar population synthesis models, initial mass functions, ionisation parameter, gas density, geometry, abundance pattern, elemental depletion, and dust—on spectral diagnostics.
Furthermore, we demonstrate the versatility of \textsc{synthesizer} through its application in different scenarios ranging from exploring  emission in toy parametric models to large-volume cosmological simulations with realistic star formation and metal enrichment histories. Taken together, \textsc{synthesizer} provides a flexible, physically motivated framework to model stellar and nebular emissions, serving as a vital link between theory and observations in the era of next-generation spectroscopic missions.

\end{abstract}

\maketitle

\section{Introduction}\label{sec:intro}

Emission from photoionised gas provides valuable insights into both the physical conditions of the ionised gas, such as its density, temperature, and chemical composition. Additionally, it provides the properties of the incoming ionising radiation field, including its intensity, spectral shape, and source characteristics. 
Analysing this emission allows us to constrain key aspects of star formation, infer the presence of active galactic nuclei (AGN), and map the interplay between radiation and the surrounding interstellar medium \citep[see review by][]{Kewley_review}.

Key to interpreting observations is photoionisation modelling, through which we can simulate the interaction between ionising radiation and the surrounding gas. 
By specifying the properties of the radiation field and the physical conditions of the gas, these models predict the resulting emission spectra, including both line and continuum emission.
Examples of commonly used models include \cloudy\ \citep{Ferland1998, Ferland2013, Ferland2017, Chatzikos2023, Gunasekera2023b} and \texttt{Mappings} \citep{Sutherland1993, Groves2004, Allen2008, Jin2022, 2025MNRAS.539..621P}.

Photoionisation modelling is now routinely combined \citep[e.g.][]{CharlotLonghetti2001, Zackrisson2001, AndersFritze2003, Gutkin2016, Byler2017, Plat2019, Newman2025} with the intrinsic spectrum of stellar populations as predicted by various stellar population synthesis (SPS) models \citep[e.g][]{BC93, Maraston1998, BC03, Maraston2005, FSPS, BPASS, Maraston2011, Maraston2020, BPASS-2.2.1,  BPASS-2.3}. 
This type of modelling has also been used to model the emission characteristics of the narrow and broad line regions near active galactic nuclei \cite[AGN, \eg][]{Feltre2016,Hirschmann2019, wu2024understandingbroadlineregionactive,Arjona2026,Zhang2026}.
By connecting the underlying physical conditions of the gas to its emergent spectrum, these models provide a crucial link between galaxy properties and observable signatures.

These frameworks can be used directly to interpret observations, for example by linking broadband photometry, and emission line strengths and ratios to the physical conditions of the gas, such as density, metallicity, or the hardness of the ionising radiation field \citep[see e.g.][]{Gutkin2016, Kewley_review,Wilkins2020}. Alternatively, they can be incorporated into spectral energy distribution (SED) fitting codes, providing a physically motivated framework for constructing model spectra that account for both stellar and nebular contributions \citep[e.g][]{Starlight1, BAGPIPES, CIGALE, Prospector}. By applying these models to observational data, one can infer key physical properties not only of the underlying stellar populations but also of the surrounding photoionised gas.

This type of modelling is also increasingly integrated into synthetic observation pipelines. These pipelines generate realistic mock imaging, photometry, line strengths, and spectra from galaxy formation models, including semi-analytic models \citep[e.g][]{Orsi2014, yung18, Baugh2022} and hydrodynamical simulations \citep[e.g][]{Wilkins2013, Barrow2017, Hirschmann2017, Ceverino2019, shen20, Wilkins2020, garg22, FLARES-II, Hirschmann2023, Katz2023,Lumen2026}. 
By generating these observables, these pipelines enable a direct comparison between simulations and observational data. This framework captures the interplay between stars, gas, dust, and AGN, providing a robust way to investigate the physical processes that shape galaxy evolution and their observable properties across cosmic time.

We have recently developed \synthesizer, introduced in \cite{synthesizer} and \cite{synthesizer-joss}, a flexible and comprehensive open source software package designed to produce synthetic observations for a wide range of galaxy models, from simple parametric representations to complex cosmological simulations. 
By integrating SPS models with photoionisation modelling via \cloudy, \synthesizer\ produces detailed spectral energy distributions (SED), emission line fluxes, and broadband photometry. 
Furthermore, it supports multiple parametric forms for the star formation and metal enrichment history, and realistic dust attenuation that incorporates line-of-sight effects.
By bridging the gap between theoretical models and observations, \synthesizer\ serves as an interface between simulations and data, generating synthetic observables that can be directly compared with real observations.

In this work, we present a detailed description of how \synthesizer\ models the emission from gas photoionised by stellar populations, including both nebular line and continuum emission. We demonstrate its versatility through a series of example applications, ranging from simple parametric galaxy models to fully synthetic observations extracted from cosmological hydrodynamical simulations. These examples illustrate how \synthesizer\ can be used to predict spectra, emission-line diagnostics, photometry, and imaging.

This article is organised as follows. In Section \ref{sec:synthesizer}, we briefly describe the relevant aspects of the \synthesizer\ code, with a particular focus on the construction of stellar grids using population synthesis models. A full description of the AGN spectral grid construction is beyond the scope of this work and will be presented in Vijayan et al. (in preparation). Section \ref{sec:modelling:photoionisation} then outlines the photoionisation modelling used to generate emission from ionised gas. In Section \ref{sec:exploration}, we examine the impact of key modelling assumptions, including the 
metallicity (\S~\ref{sec:exploration.metallicity}) and age (\S~\ref{sec:exploration.age}) of the stellar population, 
choice of stellar population synthesis model (\S~\ref{sec:exploration:sps}), 
initial mass function (\S~\ref{sec:exploration:imf}),
ionisation parameter (\S~\ref{sec:exploration:ionisation_parameter}), 
hydrogen density (\S~\ref{sec:exploration:hydrogen_density}), 
geometry (\S~\ref{sec:exploration:geometry}), 
abundance pattern (\S~\ref{sec:exploration:abundances}), and 
dust grain modelling (\S~\ref{sec:exploration:depletion}). 
Section~\ref{sec:examples} presents example applications of \synthesizer, ranging from a simple ``parametric'' galaxy, to the processing of an individual galaxy from the TNG50 simulation \cite[]{Nelson2019,Pillepich2019}. Finally, we present an application to the full EAGLE simulation to predict the redshift evolution of the hydrogen recombination line luminosity function and equivalent width distribution. We present our conclusions in Section \ref{sec:conclusions}.

\section{Synthesizer}\label{sec:synthesizer}

\textsc{Synthesizer} \citep{synthesizer, synthesizer-joss} is a fast, flexible, fully open-source and version controlled software package designed to generate a wide range of synthetic astronomical observables, including spectra (with detailed emission-line properties), photometry, and imaging. The code can be applied to simple parametric galaxy models and to outputs from state-of-the-art galaxy formation simulations, encompassing semi-analytic models as well as cosmological hydrodynamical simulations.

In this section, we provide a concise description of the core components and functionality of \synthesizer\ that are most relevant to this work. A comprehensive overview of the software, its design philosophy, and performance tests can be found in \citet{synthesizer} and \citet{synthesizer-joss}. \synthesizer\ documentation can be found at \href{https://synthesizer-project.github.io}{https://synthesizer-project.github.io}.

\subsection{Grids}\label{sec:modelling:synthesizer:grids}

At the core of \synthesizer\ are \emph{grids}. 
We use the term \emph{grid}/\texttt{Grid} to refer to both the underlying HDF5 data file and its corresponding \synthesizer\ object for manipulation (see Section 4 of \citealt{synthesizer}).
In essence, grids store spectra, and where available, the surviving mass of the simple stellar population (SSP), the specific ionising photon production rate (for H$^{+}$ and He$^{2+}$ ions), and emission line quantities predicted by photoionisation models as functions of a set of physical parameters.

Standard stellar population models are characterised by age and metallicity, but they may also incorporate additional parameters such as 
$\alpha$-enhancement \citep{BPASS-2.3} or variations in the initial mass function (IMF) \citep{FSPS}.
Grids without photoionisation modelling contain only `pure stellar' spectra, referred to in \synthesizer\ as \emph{
incident spectra}, while grids that have undergone photoionisation processing incorporate the corresponding photoionised spectra and emission-line quantities. We denote the \synthesizer\ naming for different quantities corresponding to spectra and emission-lines here for posterity:
\begin{itemize}[leftmargin=10pt]
    \item Spectra
    \begin{itemize}[leftmargin=10pt]
        \item \emph{incident}: pure stellar spectra;
        \item \emph{linecont}: the contribution of line emission to the spectra obtained from photoionisation modelling;
        \item \emph{nebular}: the nebular spectra obtained from photoionisation modelling; and
        \item \emph{transmitted}: the incident spectrum that is transmitted through the gas during photoionisation modelling.
    \end{itemize}
    \item Lines
    \begin{itemize}[leftmargin=10pt]
        \item \emph{id}: IDs of available emission lines in the grid as referenced by \cloudy\ (\eg\ H 1 6562.80A for \ha);
        \item \emph{nebular\_continuum}: the nebular continuum emission at the position (\ie\ wavelength) of the lines;
        \item \emph{transmitted}: the transmitted emission at the position of the lines;
        \item \emph{total\_continuum}: the sum of the nebular continuum emission and transmitted emission at the position of the lines.
    \end{itemize}
\end{itemize}

We have developed a companion package, \texttt{syncretize}\footnote{\href{https://github.com/synthesizer-project/syncretize}{https://github.com/synthesizer-project/syncretize}}, to streamline and automate grid construction for \synthesizer. 
The package hosts pipelines for constructing both pure-stellar (incident) grids and grids incorporating photoionisation models. Incident grids are always generated at the native spectral resolution of the underlying stellar population synthesis (SPS) model. This ensures that the full information content of the original spectra is preserved prior to any subsequent rebinning or convolution.

In line with the \synthesizer\ philosophy of openness and interoperability, we have converted a diverse suite of SPS models into \synthesizer\ grids. 
These include a growing library of SPS models that currently includes,
Binary Population and Spectral Synthesis models \citep[BPASS; versions 2.2.1 and 2.3;][]{BPASS, BPASS-2.2.1, BPASS-2.3}, the \citet{BC03} models and their 2016 revision (encompassing multiple parameter configurations), the ``Maraston'' models \citep{Maraston2011, Maraston2020, Newman2025}, the Flexible Stellar Population Synthesis framework \citep[FSPS;][]{FSPS}, and Yggdrasil \cite[]{Yggdrasil2011}. We refer the reader to Section~4.1 in \cite{synthesizer} and \href{https://synthesizer-project.github.io/synthesizer/emission_grids/precomputed_grids/sps_grids.html}{SPS Grids} in the \synthesizer\ documentation for the grid naming convention. 
The link to the full collection of \synthesizer\ SPS grids is available on the documentation page.

Each grid is provided for a range of IMFs and stellar libraries, depending on the capabilities of the underlying SPS model. For example, BPASS v2.2.1 includes several discrete IMF variants, while FSPS offers extensive flexibility, allowing users to vary both the slopes and the low- and high-mass cut-offs of the IMF. 
Unless stated otherwise, we adopt the BPASS v2.2.1 \citep{BPASS-2.2.1} binary variant as our default SPS model to showcase the capabilities of \synthesizer, assuming a \citet{ChabrierIMF} IMF over the range $0.1$–$300\, \Msun$. In Sections~\ref{sec:exploration:sps}, \ref{sec:exploration:imf}, \ref{sec:examples:haew}, and \ref{sec:examples:eagle}, we briefly explore the effects of varying the SPS model and the IMF, respectively.

\subsection{Particle and parametric emitters}
\synthesizer\ also supports a range of particle- and parametric-based approaches for generating synthetic observations. In the particle approach, outputs from hydrodynamical simulations are used directly by mapping individual particles or resolution elements to spectra. In contrast, the parametric approach constructs models from analytic parametrisations of the star formation and metal enrichment histories (SFZHs), analogous to approaches commonly used in SED fitting. This enables flexible exploration of the underlying physical parameter space.

\subsection{Emission models}\label{sec:synthesizer.emodel}
\synthesizer\ relies on emission models to produce observables such as spectra, lines, photometry, or imaging.
In \synthesizer, emission models are networks of individual steps dictating how the emissions from a component are processed before reaching an observer.
Each galaxy component can be associated with an emission model, which determines its predicted emission. This is frequently done, though not exclusively, by using a pre-computed grid of stellar or nebular properties.

\synthesizer\ provides several built-in emission model options for stellar populations. The simplest, \verb|IncidentEmission|, assigns to each element its incident (pure stellar) spectrum based solely on age and metallicity, without accounting for any reprocessing by the surrounding gas or dust.  

More sophisticated models incorporate the effects of gas and dust on stellar emission. For example, the \verb|NebularEmission| model predicts nebular spectra and line luminosities by associating each element with an individual \hii\ region. 
Compared to \verb|IncidentEmission|, \verb|NebularEmission| introduces two additional free parameters: the escape fraction of ionising photons, $f_{\rm esc}$, and the escape fraction of Lyman-$\alpha$ emission, $f_{\rm Ly-\alpha,esc}$. Depending on the chosen grid, additional parameters may need to be specified, including the ionisation parameter, hydrogen density, or 
scaling of metal abundance ratios.

An extension of this approach is the \verb|IntrinsicEmission| model, which not only calculates nebular emission but also tracks the escaping pure stellar light and the emission transmitted through the gas itself, described by the \verb|TransmittedEmission| model. Like \verb|NebularEmission|, it depends on $f_{\rm esc}$ and $f_{\rm Ly\text{-}\alpha, esc}$, along with any grid-specific parameters. In \S\ref{sec:examples:parametric}, we illustrate an example of combining a simple parametric galaxy with this emission model to predict nebular emission. \synthesizer\ also includes emission models that account for galaxy dust attenuation and its re-emission, which are employed in some of the more complex examples presented in Section~\ref{sec:examples}.

For a comprehensive description of various emission models provided by \synthesizer, we refer the reader to section~5 of \cite{synthesizer} and \href{https://synthesizer-project.github.io/synthesizer/emission_models/emission_models.html}{Emission Models} in the \synthesizer\ documentation. The photoionisation modelling included in these emission models is described in Section \ref{sec:modelling:photoionisation}.

\subsection{Emissions (SEDs, photometry, and emission lines)}
\synthesizer\ couples emission models with various components to generate a range of outputs,
including spectra, photometry, and emission-line quantities - so called \emph{emissions}. These are encapsulated within a set of dedicated classes: \verb|emissions.sed.Sed| for spectra, \verb|emissions.PhotometryCollection| for photometric data, and \verb|emissions.lines.LineCollection| for collections of emission lines. Emission data can also be extracted directly from a grid object, a functionality demonstrated in \S\ref{sec:examples:grid}.

\verb|SED| objects store spectral data, while \verb|LineCollection| objects contain both line and continuum luminosities as well as equivalent widths. Each emission class provides a suite of utility methods. For instance, the \verb|LineCollection| class includes tools for computing observable line fluxes, line ratios, and diagnostic diagrams, along with a range of pre-defined ratio and diagnostic definitions. Similarly, the \verb|Sed| class provides methods for deriving broadband fluxes or luminosities, and spectral indices, with several built-in definitions such as the UV continuum slope ($\beta$) and the Balmer break, used in the exploration section (Section~\ref{sec:exploration}).

\section{Photoionisation Modelling}\label{sec:modelling:photoionisation}
Photoionisation modelling of SPS grids in \synthesizer\ is handled by the dedicated open source package \texttt{syncretize},
part of the broader \synthesizer\ project. Our approach is similar to established methodologies in literature \citep[e.g;][]{Gutkin2016, Byler2017, Wilkins2020, Newman2025}.
In brief, each grid point in an incident stellar population grid is processed using a photoionisation code. Currently, \synthesizer\ only supports \cloudy\ \citep{Ferland1998, Ferland2013, Ferland2017} versions from 2017 onwards, with the $2023$ version \citep[C23.01;][]{Chatzikos2023, Gunasekera2023b} as the default.
In future releases, we plan to incorporate additional photoionisation codes, such as \texttt{Mappings} \citep[e.g.][]{Sutherland1993, Jin2022}.

In most cases, we adopt a single set of photoionisation modelling assumptions. In this cases, the dimensionality of the incident grid remains unchanged, but additional quantities — described in \S\ref{sec:modelling:photoionisation:outputs} — are associated with each grid point. However, we have also generated grids in which key photoionisation parameters, such as the hydrogen density and ionisation parameter, are varied. This increases the dimensionality of the grid, and when multiple parameters are varied simultaneously, the resulting grids can become large and unwieldy. Work is underway to reduce the size of these grids using emulator-based approaches, which will be seamlessly incorporated into a future release of \synthesizer.

We now describe the specific modelling choices we employ, including some of the options available within \synthesizer. In Section \ref{sec:exploration}, we explore the impact of these choices on key spectral indices, line luminosities, and line ratios.
By default, we omit CMB heating and fix the cosmic ray flux to the galactic background \cite[hydrogen cosmic ray ionisation rate of $2 \times 10^{-16}$~s$^{-1}$,][]{CR2007}, with nebular gas turbulence set to 100 km/s. 
In \cloudy, we iterate the calculation until optical depths converge (up to a maximum of 10 iterations).

\subsection{Incident Radiation Field and Geometry}\label{sec:modelling:photoionisation:incident}

The first step in our photoionisation modelling is to define both the \emph{shape} and \emph{intensity} of the incident radiation field.

\subsubsection{Shape}

The shape of the radiation field is determined by supplying \cloudy\ with the incident spectra corresponding to each grid point. This is achieved using the \verb|table SED| command, which enables \cloudy\ to read the spectral energy distribution (SED) of the stellar population element and use it as the source of ionising photons.

\begin{figure}
    \centering
    \includegraphics[width=1\columnwidth]{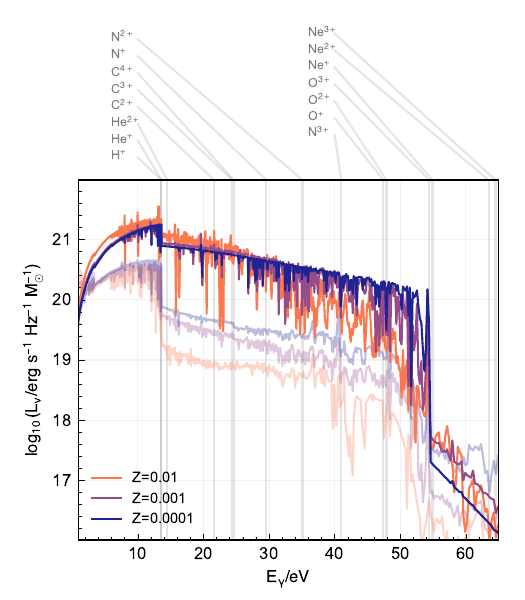}
    \caption{The Lyman-continuum spectra predicted by version 2.2.1 of BPASS assuming a \citet{ChabrierIMF} IMF with a high-mass cut-off of $300\, {\rm M_{\odot}}$ for two ages ($\log_{10}(t/{\rm yr})\in\{6, 7\}$, dark and light lines, respectively) and three metallicities ($\log_{10}(Z)\in\{-2, -3, -4\}$). The energy required to produce various ions is shown by the solid vertical lines.}
    \label{fig:modelling:incident_ionising_spectra}
\end{figure}

Figure~\ref{fig:modelling:incident_ionising_spectra} illustrates the hydrogen-ionising spectra predicted by our default SPS model and IMF. The spectra are shown for stellar populations at two ages (1 and 10 Myr) and three metallicities ($Z=0.01$, $0.001$, $0.0001$), thereby highlighting the strong dependence of the ionising output on both stellar evolution and chemical composition. 

\begin{figure}
    \centering
    \includegraphics[width=1\columnwidth]{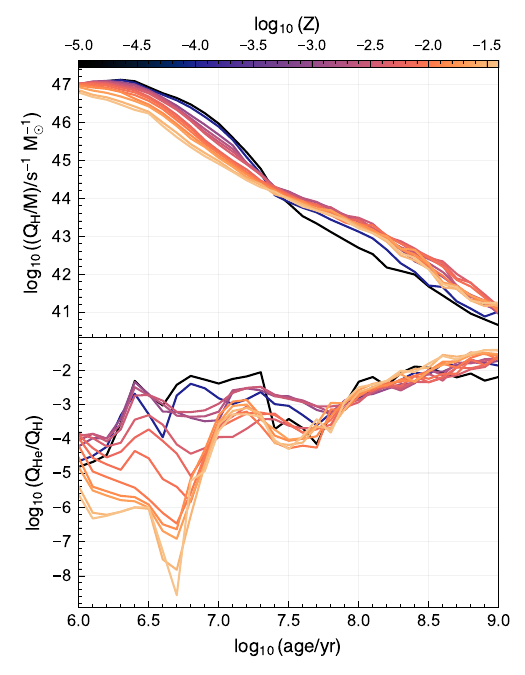}
    \caption{The specific hydrogen ionising photon production rate, $\rm Q_{H}/M$, (top-panel), and the ratio between fully-ionized helium and hydrogen ionising photon production rates, $\rm Q_{He}/Q_{H}$, (i.e. hardness, bottom-panel) as a function of age for different metallicities.}
    \label{fig:modelling:specific_ionising_luminosity}
\end{figure}

Figure~\ref{fig:modelling:specific_ionising_luminosity} presents the evolution of the specific (per unit stellar mass formed) hydrogen-ionising photon production rate (top panel) and the spectral hardness (bottom panel), defined as the ratio of helium-ionising ($>54.4$\,eV, fully ionised) to hydrogen-ionising ($>13.6$\,eV) photon production rate, as a function of stellar population age for a range of metallicities. At early times ($t \lesssim 3$ Myr), the hydrogen-ionising luminosity reaches its peak, dominated by the most massive O-type stars, before declining rapidly as these stars exhaust their fuel and leave the main sequence.  Metallicity sets both the normalisation and the persistence of this output. Lower-metallicity populations maintain higher ionising luminosities and harder spectra for longer, while higher-metallicity populations fade and soften quicker due to stronger line-driven winds and cooler stellar atmospheres. 

\subsubsection{Intensity and Geometry}
The intensity of the incident radiation provided to \cloudy\ is dependent on the assumed geometry of the gas —  specifically, whether it is spherical or plane-parallel.
The distinction is determined by the relative size of the cloud’s inner radius, $R_{0}$, compared to the Str{\"o}mgren radius, $R_{S}$:
\begin{equation}\label{eq:stromgren_radius}
R_{S} = \sqrt[3]{\frac{3Q_{\rm H}}{4\pi n_{\rm H}^2\epsilon \alpha_{B}}},
\end{equation}
where $Q_{\rm H}$ is the hydrogen ionising photon production rate, $n_{\rm H}$ is the hydrogen number density, $\epsilon$ is the volume filling factor (we fix this to 1), and $\alpha_{B}$ is the case B recombination coefficient. 

For a plane-parallel geometry, the intensity of the incident radiation can be conveniently specified using the ionisation parameter, $U$, defined as 
\begin{equation}\label{eq:ionisation_parameter}
    U = \frac{Q_{\rm H}}{4 \pi R^2 n_{\rm H} c}.
\end{equation}
The quantity is usually evaluated at the inner face ($R_{0}$) of the cloud, $U_0$. It effectively sets the strength of the radiation field relative to the gas density and therefore determines the ionisation structure and emission properties of the cloud. In \cloudy, $U_0$ can be set directly using the command \verb|ionization parameter|, which automatically scales the radiation field for the chosen geometry.

For a spherical geometry, it is more natural to define the intensity using the ionisation parameter at the Str{\"o}mgren sphere radius, $U_S$, which can be expressed as
\begin{equation}\label{eqn:stromgren_ionisation_parameter}
U_{S} \approx \frac{\langle U \rangle}{3},
\end{equation}
where $\langle U \rangle$ is the ionisation parameter averaged over the volume of the Str{\"o}mgren sphere. This approach provides a more physically meaningful measure of the radiation field relative to the gas in a fully spherical configuration.

Since \cloudy\ does not allow $U_S$ to be set directly, we instead specify the corresponding ionising photon production rate, $Q_{\rm H}$, using Equation~\ref{eq:ionisation_parameter} to achieve the desired volume-averaged ionisation parameter. In \cloudy, spherical geometry must be initialised using the \verb|sphere| command and by setting the inner radius, $r_{\rm in}$. By default, we adopt $R_{0}=0.01$ pc, set using \verb|radius -2 log parsecs|. This choice ensures that $R_{0} \ll R_S$, so the geometry behaves effectively as a true sphere for the range of intensities considered in our models.

\subsubsection{Ionisation Parameter}

When setting $U_S$ or $U_0$ for different models—e.g., varying age, metallicity, or other stellar population properties—there are two possible approaches. The first is to fix $U_S$ or $U_0$ for all models using typical values in the range $[0.0001, 0.1]$. However, as shown in Figure~\ref{fig:modelling:specific_ionising_luminosity}, the ionising photon production rate evolves rapidly with stellar population age and can vary significantly with metallicity. Keeping the ionisation parameter (and hydrogen density) fixed under these conditions effectively imposes an evolution in the geometry of the system.

An alternative approach is to adopt a \emph{reference} ionisation parameter and scale the ionisation parameter at each grid point according to the value of $Q_{\rm H}$ relative to a reference grid point. For plane-parallel geometry, $U_0$ scales directly with the hydrogen ionising photon production rate, $Q_{\rm H}(t, Z, \dots)$. For spherical geometry, $U_S$ scales as the cube root of the ionising photon production rate:
\begin{equation}\label{eqn:reference_U}
U_S(t, Z, \dots) = \left(\frac{Q_{\rm H}(t, Z, \dots)}{Q_{\rm H, ref}}\right)^{1/3} U_{\rm ref},
\end{equation}
where $Q_{\rm H, ref}$ and $U_{\rm ref}$ are the reference hydrogen photon production rate and ionisation parameter, respectively. 

\begin{figure}
    \centering
    \includegraphics[width=1\columnwidth]{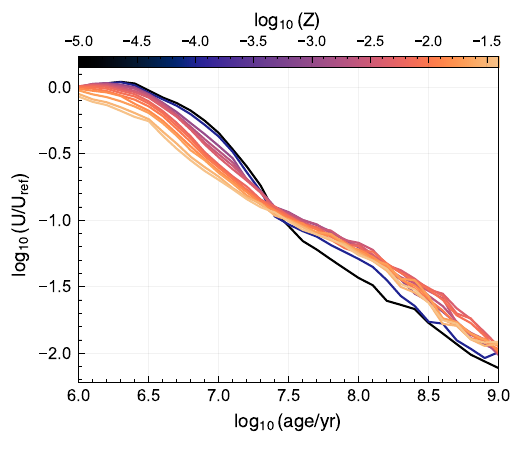}
    \caption{The ionisation parameter, $U$, relative to the default reference value $U_{\rm ref}$, as a function of age and metallicity for our default SPS model and IMF.}
    \label{fig:modelling:ionisation_parameter}
\end{figure}

Figure~\ref{fig:modelling:ionisation_parameter} shows the ionisation parameter relative to the reference value, defined at $t = 1$ Myr and $Z = 0.01$, as a function of age and metallicity for our default SPS model and IMF. The relative ionisation parameter, $U_S / U_{\rm ref}$, varies from $0.1 \to -1.9$ dex across $\log_{10}(t/{\rm yr}) = [6, 9]$ and $\log_{10}(Z) = [-5, -1.5]$. This variation reflects the rapid evolution of the ionising photon output with stellar age and its weak dependence on metallicity, highlighting the importance of scaling $U_S$ appropriately when modelling H\textsc{ii} regions around stellar populations of different ages and compositions.

By default, we adopt a spherical geometry and use the reference ionisation parameter model, with $U_{\rm ref} = 0.01$ defined at $t = 1$ Myr and $Z = 0.01$. In \S\ref{sec:exploration:ionisation_parameter}, we explore the effects of alternative geometries as well as different choices for scaling the ionisation parameter.

\subsection{Density}
The emissivity of metal lines, primarily arising from collisional excitation, is roughly proportional to the gas density squared in the low-density regime, while transitioning to a linear dependence above its critical density \cite[\eg][]{Draine_physics_of_ISGM}.
Thus, it is important to specify the density structure of the gas. By default, we adopt a hydrogen number density of $10^{2.5}\ {\rm cm^{-3}}$ \cite[similar to values adopted in SED fitting and forward modelling works, \eg][]{BAGPIPES,Wilkins2020,Lumen2026}, set using the \verb|hden <value>| command, and assume a constant density throughout the cloud via the \verb|constant density| command. The effects of varying the hydrogen density on emission lines and spectra are explored in \S\ref{sec:exploration:hydrogen_density}.

\subsection{Composition}\label{sec:modelling:photoionisation:abundances}
\synthesizer\ provides a dedicated module, \verb|synthesizer.abundances|, to specify the composition of the surrounding gas, including the gas-phase abundances of different elements and the fraction locked up in dust grains.
In practice this begins by defining a reference abundance pattern, any additional scaling, a depletion pattern, and then setting the metallicity. 
By default, we set the gas-phase metallicity (including dust depletion) equal to the metallicity of the SSP in our calculations. 
It should be noted, however, that this assumption may not be fully self-consistent with the diverse elemental abundances provided by the various SPS models employed in \synthesizer.

\subsubsection{Reference abundance pattern}

The reference abundance pattern specifies the total elemental abundances at a given reference metallicity. Consistent with the \synthesizer\ philosophy, several options are available. These include the popular solar abundance pattern of \citet{Asplund2009}, the pattern adopted by \citet{Gutkin2016} in their modelling, and the Galactic Concordance model of \citet{Nicholls2017}. By default, we adopt the Galactic Concordance model since this also includes additional metallicity dependent scalings; however, the differences between this and the other patterns are negligible at Solar metallicity \cite[$Z_\odot = 0.0134$,][]{Asplund2009}. 

\subsubsection{Abundance scaling}

\begin{figure}
    \centering
    \includegraphics[width=1\columnwidth]{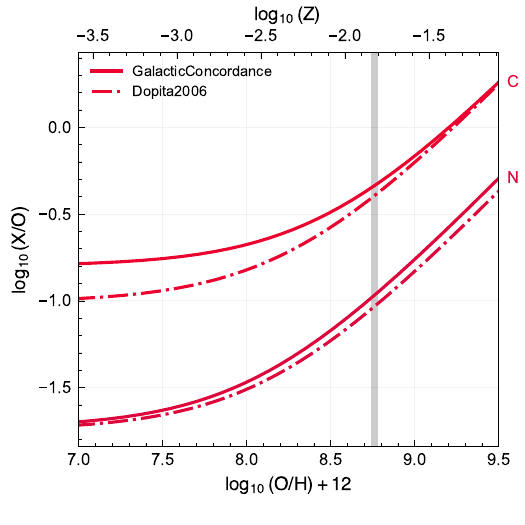}
    \caption{The abundance scaling of Carbon and Nitrogen relative to Oxygen for the \protect\cite{Dopita2006} and Galactic Concordance (default) models in \synthesizer. The vertical grey line marks the solar metallicity ($Z=0.0134$ or $12+{\rm log}_{10}({\rm O/H})=8.69$).}
    \label{fig:modelling:abundance_scalings}
\end{figure}

By default, \synthesizer\ scales the chosen reference abundance pattern uniformly with metallicity. However, it is well established that not all elements vary in this simple way — for instance, secondary production processes play a key role for elements such as carbon and nitrogen. To account for this, \synthesizer\ provides several options. First, in addition to specifying a reference abundance pattern and metallicity, users can explicitly set the abundances of selected elements, relative to either hydrogen or oxygen. This is particularly useful when exploring a broader grid of abundance patterns rather than varying metallicity alone. Second, \synthesizer\ includes two commonly used empirical scalings. These are the prescriptions of \citet{Dopita2006} for carbon and nitrogen, and the Galactic Concordance scalings \citep{Nicholls2017}, which cover a wider range of elements. Within the Galactic Concordance framework, users may also choose which elements the scalings are applied to. Figure \ref{fig:modelling:abundance_scalings} illustrates the behaviour of carbon and nitrogen relative to oxygen as a function of metallicity in these two models. In both cases the abundances of both carbon and nitrogen decrease more rapidly with metallicity than a simple scaling. The two models produce a similar overall trend, however, there are some differences: while the nitrogen scalings remain broadly consistent across the full metallicity range, the carbon trends diverge by $\approx 0.2$ dex at low metallicity. 
By default, we assume the Galactic Concordance scaling.

In \S\ref{sec:exploration:abundances} we examine the impact of varying these abundance scalings on emission lines and spectra.

\subsubsection{Depletion}\label{sec:modelling:photoionisation:depletion}
It is also necessary to account for the depletion of elements onto dust grains. Dust depletion removes atoms from the gas phase and incorporates them into dust grains, where they affect a variety of physical processes (see \S~\ref{sec:modelling:photoionisation:grains}). The gas-phase abundance of an element $X$ can be expressed as
\begin{align}
\left(\frac{X}{\mathrm{H}}\right)_{\mathrm{gas}}
= D_X \, \left(\frac{X}{\mathrm{H}}\right)_{\mathrm{total}},
\end{align}
where $D_X$ denotes the depletion factor for element $X$.

\synthesizer\ provides flexible machinery for incorporating dust depletion in \hii\ regions. 
An arbitrary depletion pattern may be specified directly, but several built-in options are also available through the \verb|synthesizer.abundances.depletion_models| module. These include the default depletion pattern implemented in \cloudy\ (\texttt{CloudyClassic}, see Table 7.8 in \emph{Hazy 1}), the variant adopted by \citet[][their Table 1]{Gutkin2016}, and the model of \citet{Jenkins2009} \cite[implemented in \cloudy\ as described in][]{Gunasekera2022}. Both the \texttt{CloudyClassic} and \citet{Gutkin2016} patterns can, in principle, be rescaled to yield an arbitrary dust-to-metal ratio (up to the limit where all refractory elements are fully depleted), although we do not explore this option here. By contrast, the updated \citet{Jenkins2009} prescription offers greater flexibility by parametrising the depletion pattern with a free parameter $F_{\star}$, which ranges from 0 to 1. \citet{Jenkins2009} adopt a fiducial value of $F_{\star} = 0.5$. In this model, the depletion factor $D_X$ for each element is given by
\begin{align}
D_X = 10^{B_X + A_X (F_{\star} - z_X)},
\end{align}
where $A_X$, $B_X$, and $z_X$ are empirically determined coefficients. 
Elements with higher $D_X$ values are less strongly depleted, while those with lower $D_X$ are more efficiently incorporated into dust grains. Note that F$_{\star}=0$ does not imply zero depletion.

\begin{figure}
    \centering
    \includegraphics[width=1\columnwidth]{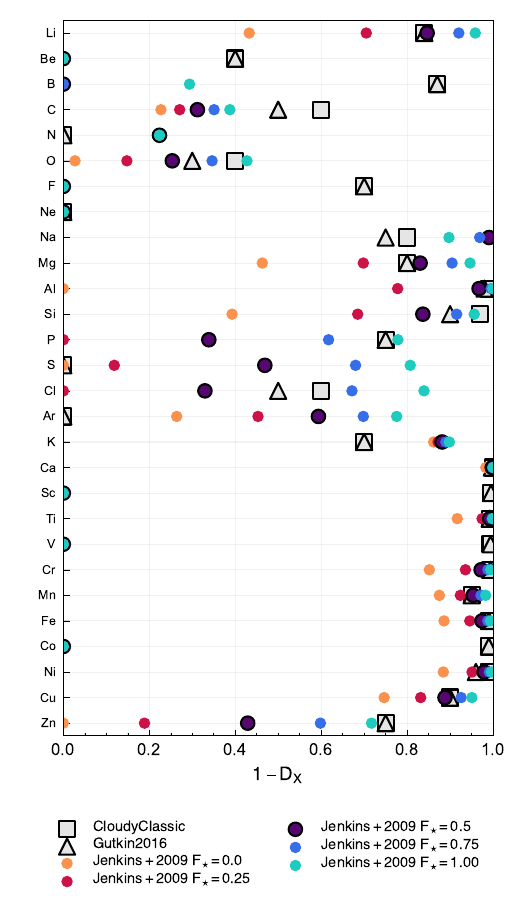}
    \caption{The depletion pattern for different elements in the \texttt{CloudyClassic}, \cite{Gutkin2016}, and \protect\cite{Jenkins2009} (F$_{\star}=0.0, 0.25, 0.5, 0.75, \, {\rm and}\, 1.0$) models available in \synthesizer. The default model choice is \protect\cite{Jenkins2009} for F$_{\star}=0.5$. Note that higher value on the x-axis imply stronger depletion.}
    \label{fig:modelling:depletion_patterns}
\end{figure}
\begin{figure}
    \centering
    \includegraphics[width=1\columnwidth]{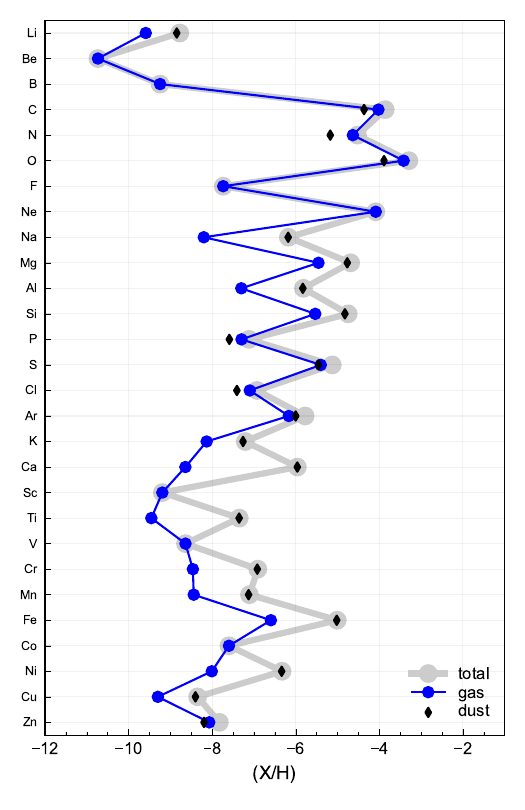}
    \caption{Total, gas, and dust phase abundances at a metallicity of $0.01$ for the \cite{Jenkins2009} depletion model with F$_{\star}=0.5$. Dust abundances for metals with zero depletion (e.g., Be, B, F) are not shown.}
    \label{fig:modelling:abundances_by_type}
\end{figure}
\begin{figure}
    \centering
    \includegraphics[width=1\columnwidth]{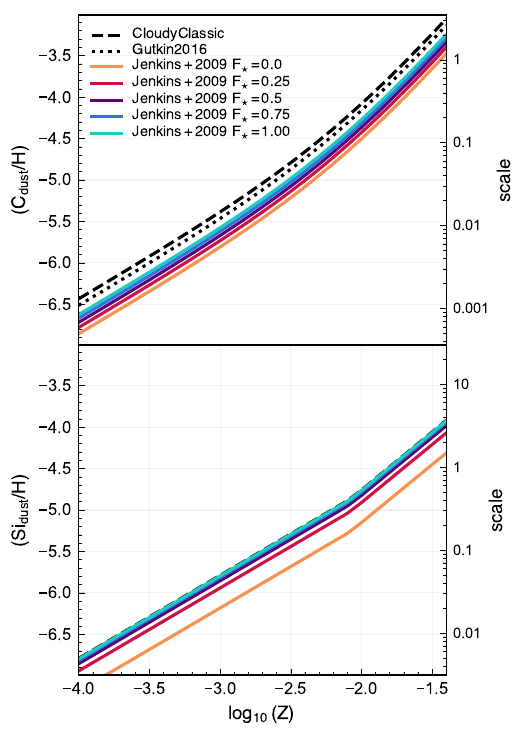}
    \caption{The abundances of carbon (top-panel) and silicon (bottom-panel) in dust as a function of metallicity for the different depletion patterns explored in this work. The right-hand side axis shows the scale parameter passed to \cloudy\ to scale the graphite and silicate grains. Here we assume $100\%$ of carbon dust is in graphite type grains.}
    \label{fig:modelling:dust_abundance}
\end{figure}

Figure~\ref{fig:modelling:depletion_patterns} compares these patterns, showing the updated \citet{Jenkins2009} model evaluated at $F_{\star}\in\{0.0,0.25,0.5,0.75,1.0\}$. Clear differences emerge between the prescriptions: for example, both \texttt{CloudyClassic} and \citet{Gutkin2016} prescriptions predict that more than 40\% of carbon is depleted, whereas the \citet{Jenkins2009} model yields a lower depletion of $\sim20$--40\%, depending on the adopted $F_{\star}$. 
By default, we adopt the \citet{Jenkins2009} model with $F_{\star}=0.5$. However, in \S\ref{sec:exploration:depletion} we examine the effects of varying $F_{\star}$  within $[0,1]$, a dust-free scenario, and the \texttt{CloudyClassic} and \citet{Gutkin2016} models.


By combining a selected depletion pattern with an abundance pattern, we can determine the fraction of each element incorporated into dust grains, and consequently, the remaining abundance in the gas phase. Figure \ref{fig:modelling:abundances_by_type} illustrates the total (dust + gas), dust, and gas-phase abundances for our default reference abundance pattern, abundance scaling, and depletion pattern at a metallicity of $0.01$.

Figure~\ref{fig:modelling:dust_abundance} shows the predicted abundances of carbon and silicon locked in dust as a function of metallicity resulting from these choices. These abundances are important because they are used to scale the amount of dust grains included in the model (see \S~\ref{sec:modelling:photoionisation:grains}) and to calculate the dust-to-metal ratio implied by each pattern. For example, combining the \texttt{CloudyClassic} depletion pattern with the Galactic Concordance abundance model at $Z=0.01$ yields a dust-to-metal ratio of $0.41$, while the \citet{Gutkin2016} pattern gives $0.35$. The \citet{Jenkins2009} prescription is more flexible, producing ratios in the range $\approx 0.14$--$0.44$ for $F_{\star} \in [0,1]$.

Although \cloudy\ provides an option for automatic element depletion, in our modelling we instead apply depletions manually by specifying the gas-phase (depleted) abundances directly in \textsc{Cloudy}, e.g. \verb|abundance <element> <gas-phase abundance> |\\ \verb|no grains|. The \verb|no grains| flag is required because we also set the grain abundances explicitly, as described in the following section.

\subsubsection{Grains}\label{sec:modelling:photoionisation:grains}

The elements locked into dust grains do not behave passively, they affect the thermal structure of the ionised gas and provide a source of opacity impacting line luminosities and transmitted spectra. Consequently when depleting elements on to grains it is essential to include grains in the calculation.

\cloudy\ provides considerable flexibility in defining grain properties, particularly the opacities used in the photoionisation calculation (see Appendix~A of \textit{Hazy~1} of \cloudy\ documentation). These opacities are obtained by combining the intrinsic optical properties of the chosen grain material with an assumed grain size distribution. 

\cloudy\ provides multiple in-built grain types and species, including silicate grains, graphite grains, and Polycyclic Aromatic Hydrocarbons (PAHs). For our modelling efforts, we use the following in \synthesizer\, (also see Section 7.9.1 in \emph{Hazy 1}):
\begin{itemize}[leftmargin=10pt]
    \item \verb|ISM| mixture: Represents the average dust found in the Milky Way ISM. This mixture is designed to reproduce observed extinction curves of the diffuse ISM and follows a power-law slope of $-3.5$ \cite[\eg][]{MRN1977} over the range grain size range, $r/{\rm \mu m} = [0.005, 0.250]$.
    \item \verb|Orion| mixture: Tailored specifically to star-forming and \hii\ regions, as found along the line of sight to stars in Orion. Similar to the \verb|ISM| mixture, it follows a $-3.5$ power-law slope, but with a truncated size range of $r/{\rm \mu m} = [0.03, 0.250]$, resulting in grains that are, on average, larger than those in the \verb|ISM| mixture.
    \item \verb|PAH| grains: These are supported via the dedicated \verb|PAH| grain type and allow for multiple customisations. In \synthesizer, we default to the \cite{Abel2008PAH} model in \cloudy, with a $-3.5$ power-law slope over the range $r/{\rm \mu m} = [0.00043, 0.0011]$.
\end{itemize}

While \cloudy\ allows for arbitrary grain mixtures (\eg\ to investigate different size distribution effects), we restrict our analysis to the 
two built-in models (\verb|Orion| and \verb|ISM|, with \verb|PAH| grains) alongside a dust-free case. By default, we assume \verb|Orion|-like grains, but explore the impact of using \verb|ISM|-like grains and dust-free case in \S\ref{sec:exploration:depletion}.

The \verb|grain|, \verb|deplete|, and \verb|abundances| commands in \cloudy\ do not always interact self-consistently, especially when using complex abundance patterns that deviate from simple metallicity scaling, as is the case here.
To ensure consistency in our modelling of grain physics, we adopt the following approach: 
\begin{enumerate}
    \item Gas-phase abundances are initialised using the command:\\
    \verb|abundance <element> <depleted_abundance>| \verb|no grains|.
    \item Specific dust grains are then included via the \verb|grains| command:\\ 
    \verb|grains <grain_model> <grain_type> <scale>|.
    \begin{itemize}[leftmargin=10pt]
        \item \verb|<grain_type>| may be either silicates or carbonaceous grains
        \item \verb|<grain_model>| can be set to \verb|ISM|, \verb|Orion| or a custom grain mixture. For PAHs, this simplifies to \verb|grains PAH <scale>|.
        \item The scale factor, \verb|<scale>|, is computed by \synthesizer\ from the ratio of the gas-phase abundances of carbon and silicon relative to those assumed in the chosen \verb|<grain_model>| and scales with metallicity. Additionally, we assume that $1\%$ of carbon is in PAHs \cite[\eg][]{Abel2008PAH}.
    \end{itemize}
\end{enumerate}
Figure~\ref{fig:modelling:dust_abundance} shows the abundances of carbon and silicon in dust as a function of metallicity, with the corresponding scale factor passed to \cloudy, for the various dust depletion patterns used in this work. 
By default, we adopt a \verb|Orion| grain mixture along with \verb|PAH| grains.

It is important to note that our implementation is not fully self-consistent, since we apply depletions to a wider set of elements than those explicitly incorporated into the grain composition specified in \cloudy.

\subsection{Stopping criteria}\label{sec:modelling:photoionisation:stopping}
\begin{table*}
\centering
\caption{Default \synthesizer\ SPS photoionisation parameters. See Section~\ref{sec:modelling:photoionisation} for details. Columns show the parameter, its value (or option), and associated remarks.}
\begin{tabular}{ccl}
\hline
Parameter & Value & Remarks \\
\hline
$U_{\rm ref}$ & 0.01 & Defined at $t=1$~Myr and $Z=0.01$ for an SPS model\\
$n_{\rm H}$ & $10^{2.5}\, {\rm cm}^{-3}$ & Constant density within the cloud\\
Reference Abundance pattern & Galactic Concordance pattern & See \cite{Nicholls2017}\\
Abundance scalings & Galactic Concordance scalings & See \cite{Nicholls2017}\\
F$_{\star}$ & 0.5 & See \cite{Jenkins2009} depletion model \\
Stopping criteria & 0.01 & Electron fraction below which calculation is stopped\\
\hline
\end{tabular}
\label{tab:default_SPS}
\end{table*}

All \cloudy\ calculations require specifying a termination criteria. There are multiple options to stop the calculations based on physical conditions, for instance, 
when the gas temperature drops below a specified value (\verb|stop temperature <value>|), 
when the free electron fraction falls too low (\verb|stop efrac <value>|), or 
upon reaching a predetermined column density (\verb|stop column density <value>|).
The specific choice depends on the modelled physical process, but for simulating stellar photoionised regions, it is common practice to stop either at a fixed column density or once the electron density has dropped to a sufficiently low value.

\begin{figure}
    \centering
    \includegraphics[width=1\columnwidth]{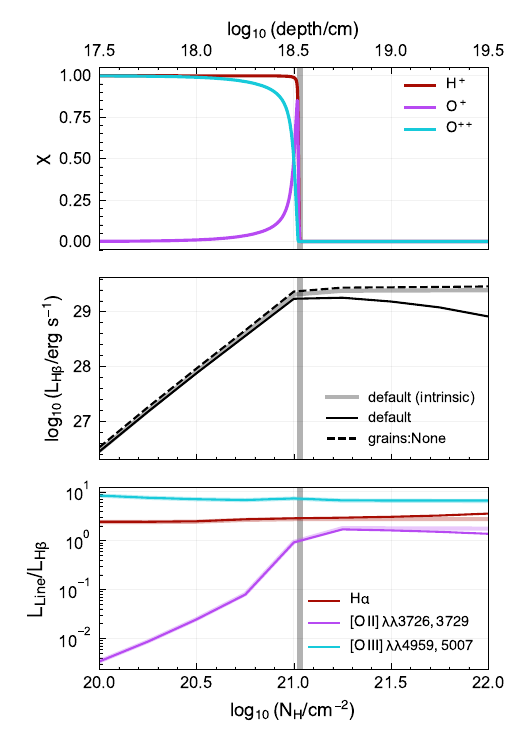}
    \caption{\emph{Top-panel} - The ionisation fraction ($\chi$) as a function of column density (and depth) for the H$^{+}$, O$^+$, and O$^{++}$ ions. \emph{Middle-panel} - The luminosity of H$\beta$ as a function of the column density for different grain assumptions. The solid vertical line in the panels denote the depth where the free electron fraction drops below $0.01$. \emph{Bottom-panel} -  The ratio of ${\rm H}\alpha$, $[{\rm O\textsc{iii}}]\lambda\lambda4959,5007$, and $[{\rm O\textsc{ii}}]\lambda\lambda3726,3729$ luminosity to the H$\beta$ luminosity as a function of column density. }
    \label{fig:modelling:column_density}
\end{figure}

To illustrate the impact of the stopping criteria, we constructed a suite of models varying column densities. All models were run with the same physical conditions: a SSP with an age of 1 Myr and metallicity, $Z=0.01$, ionisation parameter, $U_S=0.01$, and hydrogen density, $n_{H}=10^{2.5}\,{\rm cm^{-3}}$.

The top panel of Figure~\ref{fig:modelling:column_density} shows the abundance fraction of ionised hydrogen (H$^+$), singly-ionised oxygen (O$^+$), and doubly-ionised oxygen (O$^{++}$) relative to their total elemental abundance.
Hydrogen remains ionised up to a column density of $N_{H}\approx 10^{21}\ {\rm cm^{-2}}$, marking the ionisation front of the cloud. Oxygen is predominantly O$^{++}$ in the inner regions, transitioning to O$^+$ nearer the ionisation front. 

The middle panel illustrates how the \emph{intrinsic} and the \emph{emergent} H$\beta$ luminosity varies with the stopping column density. Since H$\beta$ (or H$\alpha$) luminosity directly traces the rate of hydrogen ionising photons (Q$_{\rm H}$), the luminosity increases as long as the cloud remain ionised.
Beyond the ionisation front, however, the intrinsic luminosity plateaus, while the emergent luminosity declines due to dust attenuation. We also show the effect of disabling grains. Without grains, the luminosity exceeds the intrinsic value, as no ionising photons are absorbed.

The bottom panel shows the effect of the stopping column density on three key line ratios: ${\rm H}\alpha/{\rm H}\beta$, $[{\rm O\textsc{iii}}]\lambda\lambda4959,5007 / {\rm H}\beta$, and $[{\rm O\textsc{ii}}]\lambda\lambda3726,3729 / {\rm H}\beta$. The ${\rm H}\alpha/{\rm H}\beta$ ratio remains nearly constant till the ionisation front.
While $[{\rm O\textsc{iii}}]\lambda\lambda4959,5007 / {\rm H}\beta$ remains roughly constant, $[{\rm O\textsc{ii}}]\lambda\lambda3726,3729 / {\rm H}\beta$ varies strongly through the cloud due to the low fraction of O$^+$ in the inner regions. 
Beyond the ionisation front, dust extinction affect the line luminosities. Since shorter wavelengths are more severely attenuated than H$\beta$, line ratios involving bluer lines (\eg\ $[{\rm O\textsc{ii}}]\lambda\lambda3726,3729$) decreases, and vice versa, beyond the ionisation front.
These results demonstrate that both line luminosities and line ratios are sensitive to the stopping criterion, and specifically to the column density. 

In this work, as in previous studies \citep[\eg][]{Byler2017, Wilkins2020, Newman2025}, we adopt the stopping criterion that the calculation halts once the free electron fraction falls below $0.01$, implemented via the \verb|stop efrac -2| command. This ensures the model is effectively \emph{ionisation bounded}. 
A limitation of this stopping criterion is that sources with different ionising photon luminosities penetrate to different depths. When grain physics is included, these sources experience different levels of attenuation. This effect is particularly evident when varying the ionisation parameter. A higher ionisation parameter allows radiation to penetrate more material and results in greater attenuation.

Additionally, ionising photons may escape the cloud (\emph{density bounded}) or propagate beyond the ionised region. This can be captured via an additional photoionisation grid axis describing the stopping column density. However, this increases modelling complexity and grid size, and in most cases is not justified by the resolution of 
cosmological simulations; it is therefore not included by default in our standard configuration within \texttt{syncretize}.
A more efficient approximation for photons propagating beyond the ionised region is to incorporate separate routines for dust extinction. This approach assumes that dust remains the dominant factor controlling flux loss at these large distances, which also allows us to accommodate dust optical depths derived directly from galaxy simulations. 

Table~\ref{tab:default_SPS} summarises the default SPS photoionisation model parameters adopted in \synthesizer.

\subsection{Outputs}\label{sec:modelling:photoionisation:outputs}
\cloudy\ generates a wide range of outputs, including the transmitted spectrum and the total nebular emission, the latter of which can be separated into its continuum and emission-line components. In addition, it reports the luminosities of a large number of individual emission lines.  

The \synthesizer\ \texttt{syncretize} pipeline extracts and stores a curated subset of these outputs in a new grid file. Specifically, we retain the transmitted spectra, the nebular continuum, and the nebular line spectra (see \S~\ref{sec:modelling:synthesizer:grids}). Although incident (pure-stellar) spectra are already available in separate grids, we also store the incident spectra reported directly by \cloudy. This is necessary because \cloudy\ re-samples spectra to its own energy/wavelength grid and resolution, independent of the original input spectra, thus simplifying subsequent combination of spectral components. 
In addition to spectra, we also save the luminosities of a selected subset of emission lines\footnote{Lines are selected using a relative luminosity threshold applied to both stellar and AGN models ($> -1.5$~dex with respect to \hb\ intensity); additional IR lines are included for completeness.}.  Finally, the pipeline calculates and stores continuum luminosities for the transmitted, nebular, and incident spectra, enabling straightforward computation of equivalent widths.







\section{Exploration}\label{sec:exploration}

We now examine how several key observables—including the UV continuum slope ($\beta$), emission-line luminosities, and diagnostic line ratios (as defined in Table~\ref{tab:line_ratios})—for an SSP depend on metallicity, stellar age, the form of the initial mass function, and a range of photoionisation parameters (ionisation parameter, hydrogen density, abundance pattern, elemental depletion, and the presence of dust grains). Although the influence of many of these factors has been studied extensively in the literature, we include them here to provide a consistent and comprehensive baseline for interpreting results when applying \synthesizer, for instance, to forward model cosmological simulations (see also Section \ref{sec:examples}).

\begin{table}
\centering
\caption{Emission line ratios explored in this work.}
\begin{tabular}{cl}
\hline
Ratio & Definition \\
\hline
Balmer decrement & \ha/\hb \\
R3 & $[{\rm O\textsc{iii}}]\lambda5007 / {\rm H}\beta$ \\
R2 & $[{\rm O\textsc{ii}}]\lambda\lambda3726,3729 / {\rm H}\beta$ \\
R23 & $([{\rm O\textsc{ii}}]\lambda\lambda3726,3729 + [{\rm O\textsc{iii}}]\lambda\lambda4959,5007) / {\rm H}\beta$ \\
S2 & $([\mathrm{S}\textsc{ii}]\lambda6716\ +\ [\mathrm{S}\textsc{ii}]\lambda6731)/\mathrm{H}\alpha$ \\
O32 & $[{\rm O\textsc{iii}}]\lambda5007 / [{\rm O\textsc{ii}}]\lambda\lambda3726,3729$ \\
Ne3O2 & $[{\rm Ne\textsc{iii}}]\lambda3869 / [{\rm O\textsc{ii}}]\lambda\lambda3726,3729$ \\
N2 & $[\mathrm{N}\textsc{ii}]\lambda6583/\mathrm{H}\alpha$ \\
\hline
\end{tabular}
\label{tab:line_ratios}
\end{table}

It is also important to emphasise that we focus here on only a limited set of observables. Some of the variations we explore may have only subtle effects on these quantities, while their impact on other diagnostics could be more significant. For example, except for C\textsc{iii}]$\lambda\lambda1907,1909$, none of our chosen line ratios are directly sensitive to changes in the carbon abundance, except indirectly through the role of carbon in grain physics. 

Unless otherwise stated, all predictions are based on the binary variant of the BPASS \texttt{v2.2.1} model, adopting a \citet{ChabrierIMF} IMF with an upper-mass cutoff of $300\ {\rm M_{\odot}}$. In all cases, we consider the full metallicity range available within the chosen SPS model. Except where explicitly noted in \S\ref{sec:exploration.age}, we assume an SSP age of 1~Myr.


\subsection{Metallicity}\label{sec:exploration.metallicity}

\begin{figure}
    \centering
    \includegraphics[width=\columnwidth]{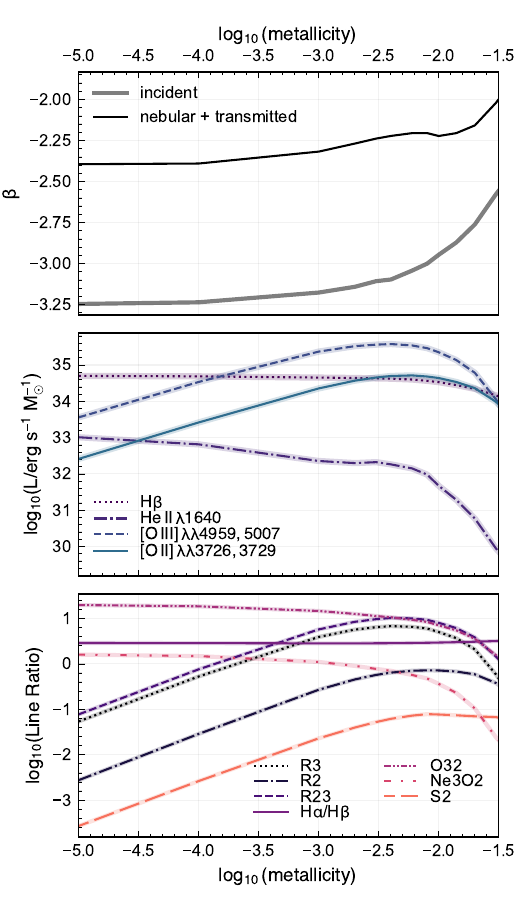}
    \caption{UV continuum slope $\beta$, line luminosities, and key line ratios as a function of metallicity for our default modelling choices.}
    \label{fig:exploration.metallicity}
\end{figure}

Figure~\ref{fig:exploration.metallicity} explores the effect of metallicity on three categories of observables: the UV continuum slope ($\beta$, top panel), selected emission-line luminosities (middle panel), and commonly used diagnostic line ratios (bottom panel) defined in Table \ref{tab:line_ratios}. 

The top panel of Figure~\ref{fig:exploration.metallicity} shows how the UV continuum slope of both the \texttt{incident} (pure stellar) and \texttt{transmitted} + \texttt{nebular} (incident spectra reprocessed by the nebula) spectra varies as a function of metallicity. 
$\beta$ measured for the \texttt{incident} spectra increases by $\approx 0.6$ over the full metallicity range (log$_{10}(Z) \in [-5.0, -1.5]$).
This reflects the fact that lower metallicity stars tend to be hotter due to lower opacities and weaker winds. Accounting for transmission through the gas, but more importantly adding nebular emission, causes the UV spectrum to significantly redden since the nebular emission is relatively flat compared to the steeply falling, bluer stellar emission. This effect is larger ($\Delta\beta\approx 0.8$) at lower metallicity due to the larger ionising luminosity.

The middle panel of Figure \ref{fig:exploration.metallicity} illustrates how the luminosities of four representative emission lines—\hb, He\textsc{ii}$\lambda1640$, \oii$\lambda\lambda3726,3729$, and \oiii$\lambda\lambda4959,5007$—vary with metallicity. Metallicity influences line emission through four primary effects: (i) changing the chemical composition of the gas, (ii) the luminosity of the ionising spectrum, (iii) hardness of the ionising spectrum (and thus the ionisation balance), and (iv) the amount of dust. 
At $Z < 0.005$, the \hb\ line luminosity remains relatively stable, but at $Z > 0.01$ it declines, driven primarily by the reduced ionising luminosity of metal-rich stellar populations, with a secondary contribution from attenuation by dust grains (see \S\ref{sec:exploration:depletion}).
The He\textsc{ii}$\lambda1640$ line declines gradually until $Z \approx 0.01$, beyond which it falls rapidly, reflecting the change in the hardness of the ionising spectrum (see Figure~\ref{fig:modelling:specific_ionising_luminosity}).
For the oxygen lines, the behaviour is more complex. At low metallicity ($Z < 0.005$), the line luminosities scale approximately with oxygen abundance. At higher metallicity, however, both \oii\ and \oiii\ exhibit a peak before declining. The decline reflects the impact of softer and less luminous ionising spectra, and more efficient cooling, which lowers the electron temperature and reduces the efficiency of collisional excitation. 
Similar to \hb, both the helium and oxygen lines are affected by secondary contribution from dust attenuation at the high-metallicity end.

The influence of metallicity on various diagnostic line ratios (R3, R2, R23, Balmer decrement (H$\alpha$/H$\beta$), O32, Ne3O2, S2) is summarised in the bottom panel of Figure~\ref{fig:exploration.metallicity}. These line ratios, except the Balmer decrement, are commonly used in combination with each other to infer galaxy metallicities.
Complementarily, Figure~\ref{fig:exploration.age} illustrates how metallicity, in combination with stellar population age, shapes the distribution in the popular \bptnii\ diagnostic diagram.

\subsection{Age}\label{sec:exploration.age}

\begin{figure}
    \centering
    \includegraphics[width=\columnwidth]{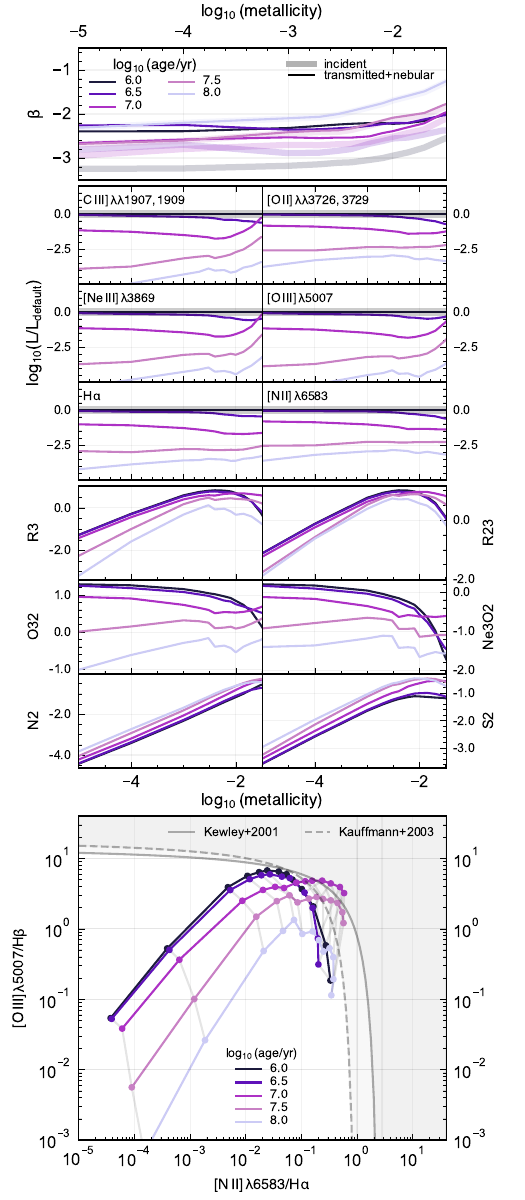}
    \caption{The impact of age on the UV continuum slope (first-panel), various line luminosities (second-panel, relative to our default model), diagnostic line ratios (third-panel), and the BPT-N\textsc{ii} diagram (bottom-panel, scatter points show different metallicities). The grey solid and dashed line are the AGN demarcation lines from \cite{Kewley2001} and \cite{Kauffmann2003}, respectively.}
    \label{fig:exploration.age}
\end{figure}

In Figure~\ref{fig:exploration.age}, we explore the effect of stellar population age, spanning $t=1$–$100\ {\rm Myr}$, on key observables as a function of metallicity.
As a stellar population ages, its most massive stars progressively leave the main sequence. The immediate consequence is a steep decline in the ionising photon luminosity once the population is older than a few Myr. As shown in Figure~\ref{fig:modelling:specific_ionising_luminosity} (top-panel), the ionising luminosity is highly dependent on both age and metallicity. 
At $Z=10^{-2}$, there is a sharp decline of roughly two orders of magnitude for ages $t=1$ to $10\ {\rm Myr}$, whereas at $Z=10^{-5}$, the decrease is milder (by $\sim 10$).
However, beyond $t=10\ {\rm Myr}$, the relationship reverses. The steepest declines are observed at low metallicities, with drops exceeding a factor of $10^3$, while higher-metallicity populations decline more moderately (by about a factor of $10^2$).
The spectral hardness also evolves strongly with age and metallicity, generally increasing with age, with the trend particularly pronounced for the high-metallicity end (see bottom-panel in Figure~\ref{fig:modelling:specific_ionising_luminosity}).
In the reference ionisation-parameter model adopted here, $U$ scales with the ionising luminosity. Thus, as populations age and $Q_{\rm H}$ decreases, the effective ionisation parameter also falls (see Figure~\ref{fig:modelling:ionisation_parameter}), which has a major influence on line ratios.

The top panel of Figure~\ref{fig:exploration.age} shows a subset of the same observables (R3, R23, O32, Ne3O2, N2, and S2) 
as Figure~\ref{fig:exploration.metallicity}, now for a wider range of stellar ages (${\rm t/yr} = 10^6, 10^{6.5}, 10^7, 10^{7.5}, {\rm and}\, 10^{8}$).
With age, the stellar continuum becomes intrinsically redder, while the contribution from nebular continuum emission steadily diminishes. By $t=100\ {\rm Myr}$, nebular emission has a negligible effect on the slope.

The second panel of Figure~\ref{fig:exploration.age} shows how key line luminosities evolve relative to our default assumption (SSP of 1 Myr age). The dominant trend is the rapid decline in ionising luminosity: hydrogen recombination lines fade by nearly four orders of magnitude between $t=1$ and $100\ {\rm Myr}$. The \oiii\ line is even more strongly affected, since its strength depends not only on the ionising photon budget but also on the ionisation parameter, which decreases with age. An important exception arises at high metallicity, where the \oiii\ luminosity declines more slowly. 
This behaviour is unique to the BPASS v2.2.1 models, in which the hardness of the ionising spectrum for metal-rich populations increases over this age range (see bottom panel of Figure~\ref{fig:modelling:specific_ionising_luminosity}).

The third panel of Figure~\ref{fig:exploration.age} shows the evolution of key emission line ratio diagnostics with age and metallicity.
Not all diagnostics respond equally to stellar age. Ratios such as R3, O32, and Ne3O2 are highly sensitive, while others—including N2, S2, and R23—show only modest changes, with N2 and S2 even rising slightly with time. 
The R3 ratio falls sharply beyond 30~Myr as the ionising spectrum weakens. However, at the highest metallicities, R3 increases between $t=3$--$10\ {\rm Myr}$, despite the overall decline in ionising photon output, owing to the spectral hardening noted above.
These trends are reflected in the bottom panel of Figure~\ref{fig:exploration.age}, which illustrates the impact of age and metallicity on the \bptnii\ diagram.

It is important to emphasise, however, that the integrated emission from star-forming galaxies is usually dominated by their youngest stellar populations, whose much higher ionising luminosities overwhelm the contribution of older stars.


\subsection{Stellar Population Synthesis model}\label{sec:exploration:sps}

One of the key strengths of the \synthesizer\ project is its flexibility, particularly in enabling the construction of model grids based on a wide range of stellar population synthesis (SPS) models and their variants. This versatility allows us to directly compare how different SPS prescriptions influence predicted observables. A complementary and more extensive analysis of this question has recently been presented by \citet{Newman2025} (see Section~4.4), who also utilised \synthesizer\ and the same default photoionisation assumptions, to which we refer the reader for further discussion.

\begin{figure}
    \centering
    \includegraphics[width=1\columnwidth]{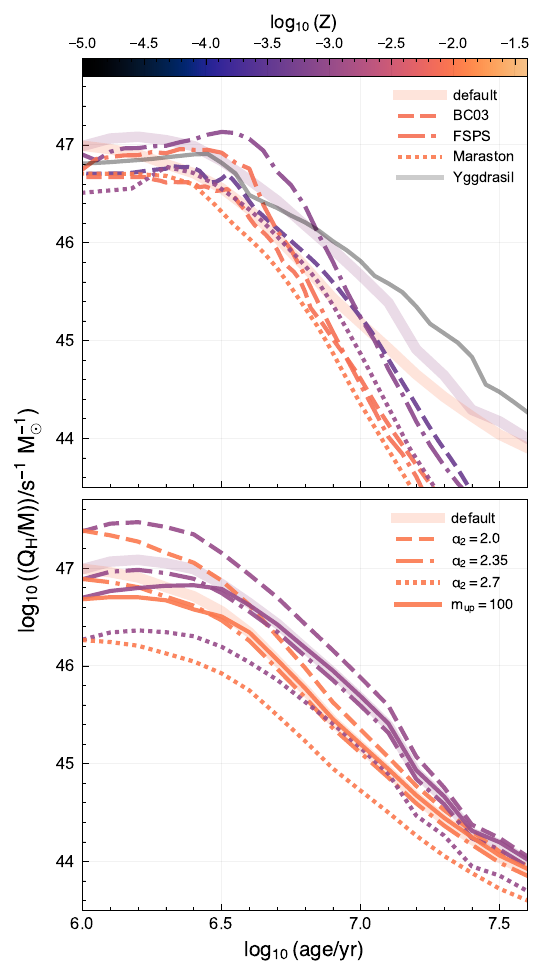}
    \caption{The impact of the choice of stellar population synthesis model (SPS, top) and initial mass function (IMF, bottom) on the specific ionising photon production rate. To maintain clarity we only show the impact at two metallicities: $Z\in\{0.001,  0.01 \}$. However, note that the Yggdrasil model is shown for a zero metallicity stellar population.}
    \label{fig:exploration:specific_ionising_luminosity_sps_imf}
\end{figure}

The top panel of Figure~\ref{fig:exploration:specific_ionising_luminosity_sps_imf} shows the evolution of the ionising photon production rate with age for \citet[][updated version from 2016]{BC03} model (with \citealt{ChabrierIMF}), FSPS \cite[][with \citealt{ChabrierIMF} IMF]{FSPS} model, and Maraston \cite[][with \citealt{kroupa} IMF]{Maraston2011} model compared to our default BPASS SPS model. We show the evolution for metallicities of $0.001$ and $0.01$ (covered by all the SPS models). 
We also show the ionising photon production rate with age for zero metallicity population III stars from the Yggdrasil SPS model \cite[]{Yggdrasil2011} for a \cite{kroupa} IMF in the $0.1-100\, \Msun$ stellar mass range.
Generally, the ionising photon production rate is higher for young and low-metallicity stellar populations. However, for $Z=0.001$, the FSPS model shows a clear peak around 3 Myr. In the case of the Maraston model, the number of ionising photons at the youngest age is lower at $Z=0.001$ compared to $Z=0.01$ \cite[see][]{Newman2025}.

\begin{figure}
    \centering
    \includegraphics[width=1\columnwidth]{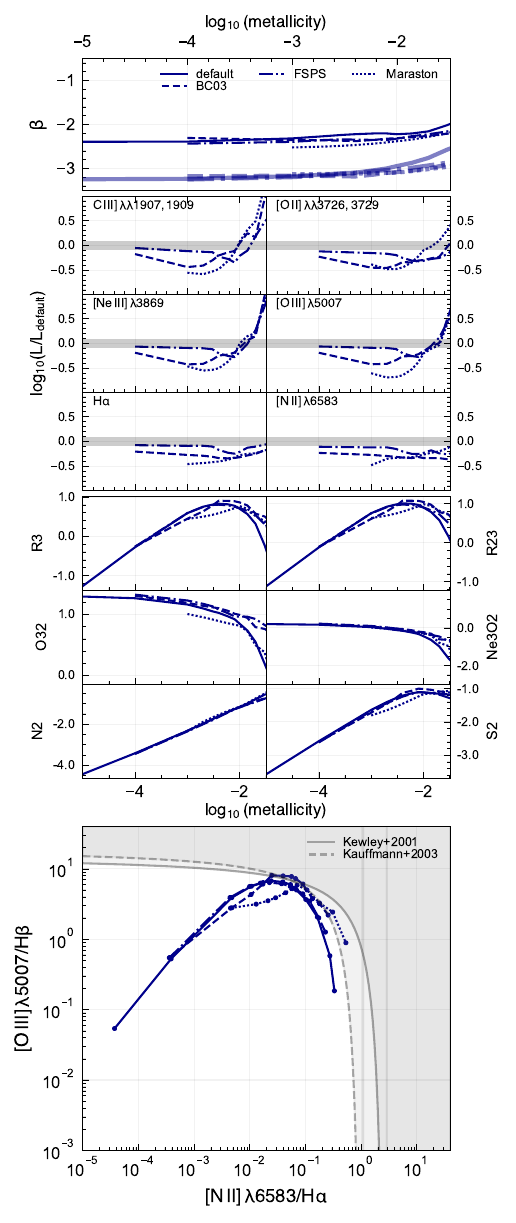}
    \caption{Same as Figure \ref{fig:exploration.age} but showing the impact of the changing the SPS model for SSP of age $10^{6}$~yr. Note that different SPS models cover distinct metallicity ranges.}
    \label{fig:exploration.sps}
\end{figure}

Figure~\ref{fig:exploration.sps} is the same as Figure~\ref{fig:exploration.age}, but now illustrating the effect of different SPS models, rather than age, on key observables. For simplicity, we only show variation with metallicity for an SSP with age $10^{6}$~yr, and omit the Yggdrasil model (since it is a zero metallicity model). The UV-continuum slope is only weakly affected, with changes only apparent at high-metallicities ($Z \approx 0.001$), implying that differences in the shape of the spectrum in the UV are small across the models. At low metallicities, the line luminosities of the default model are generally higher, reflecting the higher ionising photon rate at $10^6$~yr, as shown in Figure~\ref{fig:exploration:specific_ionising_luminosity_sps_imf} (top-panel).

The diagnostic line ratios as a function of metallicity show little variation between SPS models due to similar spectral hardness at an age of 1 Myr \cite[see also figure~13 in][for evolution with age]{Newman2025}. The higher values for certain line luminosities (\eg\ [Ne\textsc{iii}]$\lambda 3869$) or diagnostic ratios (\eg\ Ne3O2) seen at the high-metallicity end ($Z>0.02$) are due to a higher number of ionising photons for those specific ions in BC03, FSPS, or Maraston models compared to our default model 
\cite[not shown here, see Figure~14 in][]{Newman2025}, illustrating the large impact the choice of SPS model can have on key observables.

We will revisit the effect of SPS models on derived observables in \S~\ref{sec:examples}, where we explore their effect on predicted H$\alpha$ line luminosity and equivalent widths.

\subsection{Initial Mass Function}\label{sec:exploration:imf}
\begin{figure}
    \centering
    \includegraphics[width=1\columnwidth]{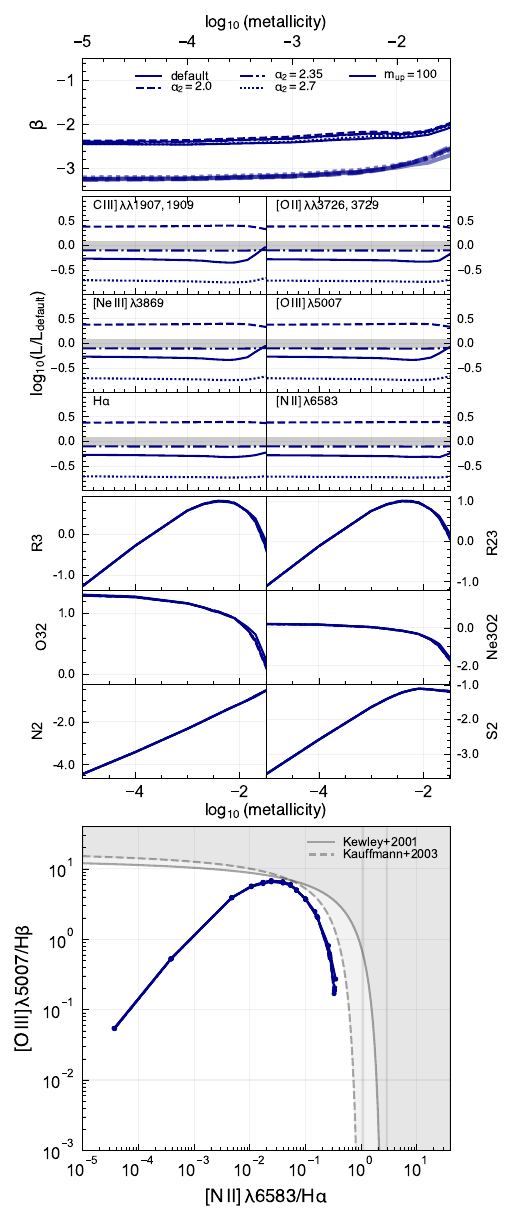}
    \caption{Same as Figure \ref{fig:exploration.sps} but showing the impact of the changes to the high-mass slope ($\alpha_2$) and cut-off of the initial mass function.}
    \label{fig:exploration.imf}
\end{figure}
We now examine how variations in the initial mass function (IMF) affect our observables, focusing on the BPASS v2.2.1 SPS model. Specifically, we explore two modifications to our fiducial choice of a \citet{ChabrierIMF} IMF. First, we replace it with a broken power-law IMF characterised by a low-mass ($0.1$--$1\ {\rm M_{\odot}}$) slope of $\alpha_1 = 1.3$ and a high-mass ($1$--$300\ {\rm M_{\odot}}$) slope of $\alpha_2 \in \{2.0, 2.35, 2.7\}$. The case $\alpha_2=2.35$ is very similar to the \citet{ChabrierIMF} IMF. Second, we lower the maximum stellar mass from the default $300\ {\rm M_{\odot}}$ to $100\ {\rm M_{\odot}}$.

Steepening the high-mass slope or reducing the maximum stellar mass decreases the number of luminous, hot, massive stars, and therefore lowers the ionising photon output. 
The bottom panel of Figure \ref{fig:exploration:specific_ionising_luminosity_sps_imf} illustrates this effect by showing how variations in the IMF impact the ionising photon luminosity at metallicities, $Z=0.001, {\rm and}\, 0.01$. At the youngest ages, steepening the slope from $2.35$ to $2.7$ reduces the ionising luminosity by $\sim0.6$ dex, while making the slope shallower, from $2.35$ to $2.0$, increases it by $\sim0.5$ dex. 
Decreasing the maximum stellar mass to $100$ M$_{\odot}$, reduces the ionising luminosity at 1 Myr by $\approx 0.4$ dex, with negligible differences by $\gtrapprox 4$ Myr.
The sensitivity to the IMF weakens with increasing metallicity and age, reflecting the combined effects of cooler stellar atmospheres and the rapid decline in the contribution from the most massive stars.

If the geometry were fixed, this would---as in the case of increasing stellar population age---directly translate into a lower ionisation parameter. However, our modelling enforces a reference ionisation parameter that is applied uniformly across different IMFs. This means that even when one IMF produces substantially more ionising photons than another (e.g. $\alpha_2=2.0$ vs.\ $\alpha_2=2.7$), the ionisation parameter remains matched at the reference point, with differences suppressed due to the cube root dependence of the changing ionising photon production rate (equation~\ref{eqn:reference_U}). In reality, a population forming with a shallower high-mass slope---and hence higher ionising luminosity---would likely sustain a higher ionisation parameter. As a result, while changes to the IMF strongly affect absolute line luminosities, they have only a modest effect on line ratios within our framework.

The influence of IMF variations on our key observables is illustrated in Figure \ref{fig:exploration.imf}. The UV continuum slope is only weakly affected, but line luminosities respond much more strongly, with changes that closely track the underlying shifts in ionising photon production. Because the ionisation parameter and spectral hardness remain largely unchanged, the resulting impact on line ratios is minimal.

We revisit the role of the IMF in \S\ref{sec:examples}, where we examine its effect on the predicted H$\alpha$ luminosity and equivalent width.

\subsection{Ionisation parameter}\label{sec:exploration:ionisation_parameter}

\begin{figure*}
    \centering
    \includegraphics[width=\columnwidth]{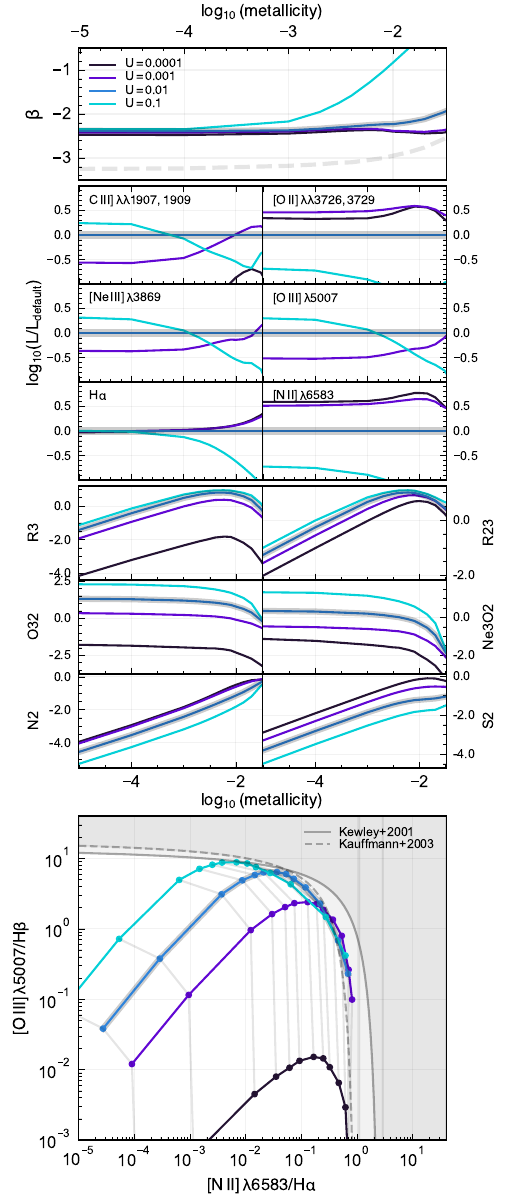} 
    \includegraphics[width=\columnwidth]{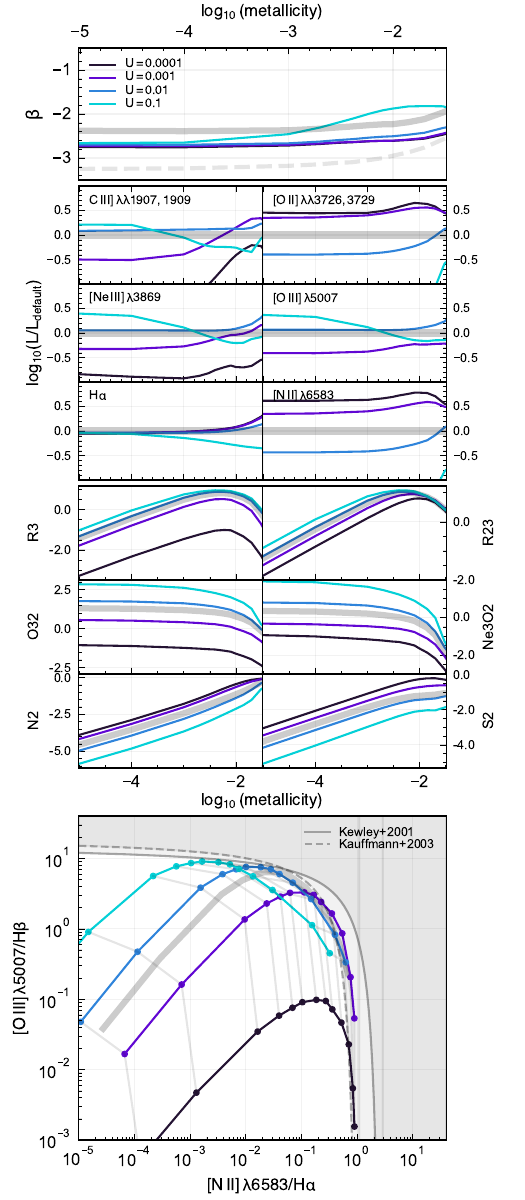} 
    \caption{Same as Figure \ref{fig:exploration.sps} but showing the effect of the choice of reference ionisation parameter (left column) and fixed ionisation parameter (right column). The default model (U$_{\rm ref}=0.01$) is also shown as the grey solid line.}
    \label{fig:exploration.ionisation_parameter}
\end{figure*}

Figure~\ref{fig:exploration.ionisation_parameter} explores the impact of the ionisation parameter, $U$, considering both the reference-scaling and fixed approaches. Like with metallicity, the effect of varying $U$ on emission line luminosities and ratios has been extensively studied in the literature \cite[\eg][]{Byler2017,Kewley_review}, we revisit this here in order to provide a consistent baseline and to place the impact of other model variations in context. 

In the left panel of Figure \ref{fig:exploration.ionisation_parameter}, we show results for models spanning $U \in {10^{-4}, 10^{-3}, 10^{-2}, 10^{-1}}$, applied consistently under both approaches. The key distinction between the two approaches lies in how the ionisation parameter is treated. In the fixed model, the ionisation parameter is held constant across all metallicities. In contrast, in the reference model it varies with metallicity, since the ionising luminosity itself depends on metallicity. As a result, for the same nominal choice of $U$, the predictions coincide at the reference point (here $Z=0.01$) but diverge at other metallicities.

In both approaches we find that the UV continuum slope reddens systematically with increasing metallicity, with the effect becoming stronger at higher ionisation parameters and most pronounced for $U=10^{-1}$. For instance, at $Z=0.01$ the slope evolves from $\beta \approx -2.4$ at $U=0.01$ to $\beta \approx -1.0$ at $U=0.1$. This behaviour is primarily driven by dust attenuation, since in our models the dust grain abundance scales with metallicity. The enhanced sensitivity to $U$ arises because higher ionisation parameters correspond to larger \hii\ regions; this requires photons to travel longer path lengths, resulting in a higher dust optical depth before the calculation terminates.
Lines at shorter wavelengths are more affected by dust compared to those at longer wavelengths. However, it should be noted that the dust optical depth of lines is also dependent on where they arise within the \hii\ region.

This behaviour also explains the strong sensitivity of the hydrogen line luminosities to changes in the ionisation parameter. In principle, and in the absence of dust, the luminosities of hydrogen recombination lines should be independent of $U$. The strong dependence we observe arises primarily from dust attenuation. 

For the metal lines, the impact of varying $U$ is more direct. Increasing $U$ shifts the ionisation balance toward higher stages---for example, enhancing the abundance of O$^{2+}$ relative to O$^{+}$---and thereby increasing the \oiii/\oii\ ratio. The strength of this dependence, however, is itself metallicity-dependent. As discussed in \S\ref{sec:exploration.metallicity}, higher metallicities lead to softer ionising spectra and more efficient cooling, both of which act to suppress the \oiii/\oii\ ratio. Analogous processes drive the behaviour of other diagnostic line ratios.

\subsection{Hydrogen density}\label{sec:exploration:hydrogen_density}
\begin{figure}
    \centering
    \includegraphics[width=1\columnwidth]{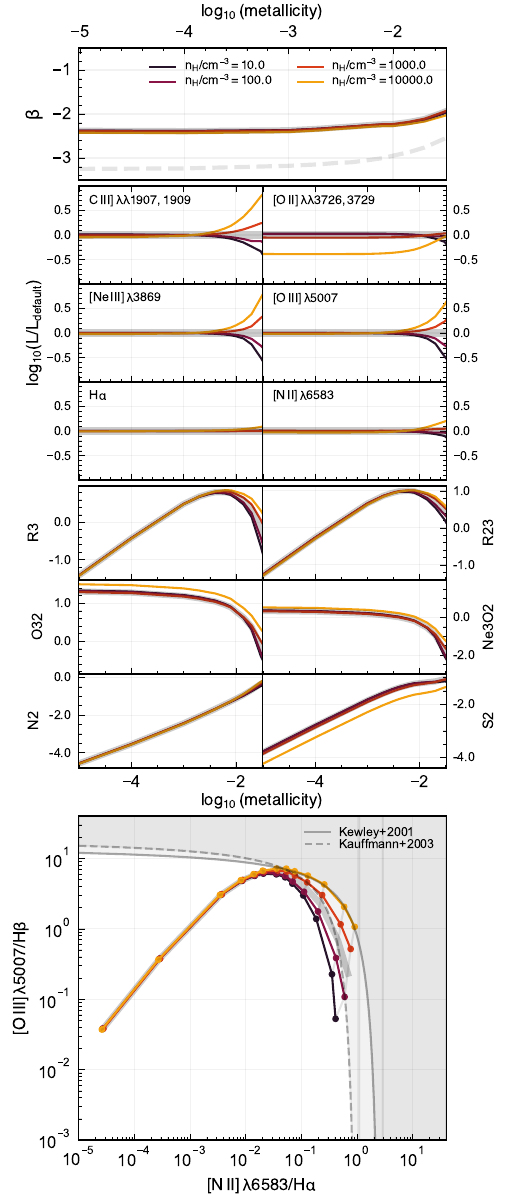}
    \caption{Same as Figure \ref{fig:exploration.sps} but showing the effect of the choice of hydrogen density $n_{\rm H}$. The default model ($n_{\rm H}=10^{2.5}\, {\rm cm}^{-3}$) is also shown as the grey solid line. }
    \label{fig:exploration:hydrogen_density}
\end{figure}

In Figure \ref{fig:exploration:hydrogen_density} we show model predictions for a range of hydrogen densities, $n_{\mathrm{H}}/{\rm cm^{-3}} \in \{10, 10^{2}, 10^{3}, 10^{4}\}$, compared to our default choice of $n_{\mathrm{H}} = 10^{2.5}\ {\rm cm^{-3}}$.  
As with the ionisation parameter, the influence of density on nebular emission has also been studied extensively in the literature \cite[\eg][]{Kewley_review,Newman2025}; our aim here is primarily to provide context for the changes introduced in subsequent sections. 

Within the explored range, the choice of hydrogen density has only a minor effect on hydrogen line luminosities, as these lines arise primarily from recombination transitions. Similarly, the impact on the nebular continuum emission—and consequently on the UV continuum slope—is small, since these are largely driven by the ionising photon flux.

In contrast, the hydrogen density plays a critical role in shaping the strengths of metal emission lines. Many metal lines originate from forbidden transitions, which are susceptible to collisional de-excitation at sufficiently high densities. 
For example, the \oii\ line has a relatively low critical density ($n_{e} \approx n_{\rm H} \sim 1.5 \times 10^3\ {\rm cm^{-3}}$), leading to noticeable sensitivity across the density range we explore. Figure~\ref{fig:exploration:hydrogen_density} shows that at low metallicities the \oii\ line strength can vary by $\sim 0.4$~dex for $n_{\rm H}=10$--$10^4\ {\rm cm^{-3}}$. The \oiii\ line, by contrast, has a much higher critical density ($n_{\rm H} \sim 7 \times 10^5\ {\rm cm^{-3}}$), well above our explored range. However, at high metallicities, the hydrogen density does influence \oiii\ emission, as efficient metal-line cooling reduces the electron temperature, which in turn affects collisional excitation rates and line luminosities.

This, in turn, affects the metallicity sequence in the \bptnii\ diagram. At high metallicity, increasing the hydrogen density enhances both the \oiii\ and [N\textsc{ii}] line luminosities, while the hydrogen recombination lines remain essentially unchanged. As a result, the high-metallicity locus shifts upward and to the right, moving it closer to the classical AGN region of the diagram.

\subsection{Geometry}\label{sec:exploration:geometry}

As noted, within \cloudy\ and the \synthesizer\ implementation, there is the ability to define the geometry where the options are the default spherical geometry and a plane parallel geometry. 

\begin{figure}
    \centering
    \includegraphics[width=\columnwidth]{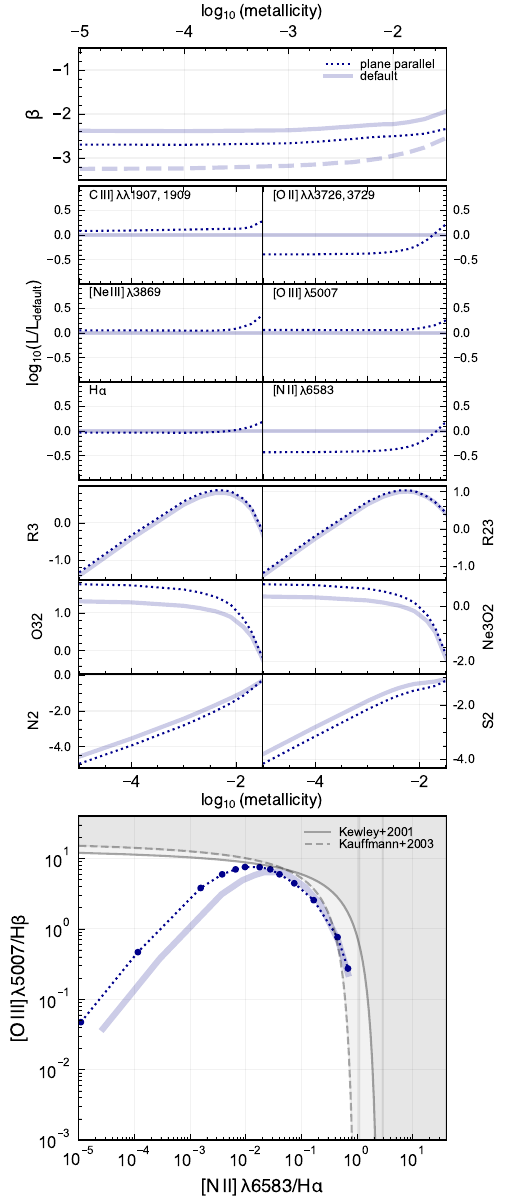}
    \caption{Same as Figure \ref{fig:exploration.sps} but showing the effect of the choice of geometry.}
    \label{fig:exploration.geometry}
\end{figure}

In Figure~\ref{fig:exploration.geometry} we illustrate the impact of adopting a plane-parallel geometry. Relative to our default spherical case, this assumption can in some instances lead to noticeable differences in the predicted line ratios and continua (see also Section~9 in \emph{Hazy 1}). Much of this discrepancy arises from the way the ionisation parameter is defined in the two geometries: in the plane-parallel case it is specified at the illuminated face of the cloud, whereas in the spherical case it is effectively averaged across the entire Str{\"o}mgren sphere. Thus, the ionisation parameter is higher in the former than the latter, which is reflected in the O32 ratio being generally higher for the plane-parallel case.

\subsection{Abundance Pattern}\label{sec:exploration:abundances}
\begin{figure}
    \centering
    \includegraphics[width=\columnwidth]{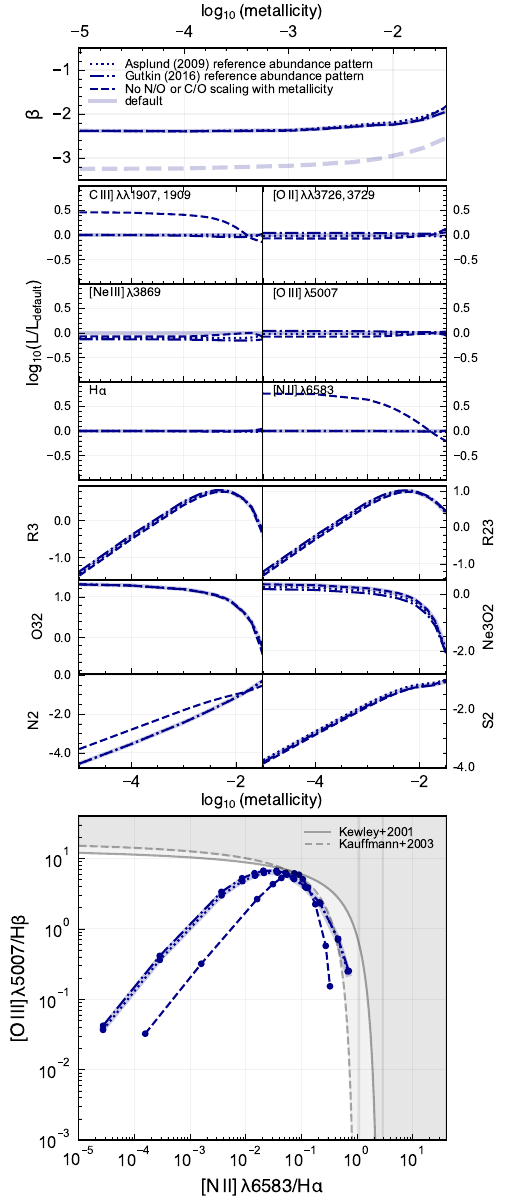} 
    \caption{Same as Figure \ref{fig:exploration.sps} but showing the effect of the choice of reference abundance pattern and scaling. Note that our default choice is the Galactic Concordance model (see \S~\ref{sec:modelling:photoionisation:abundances}).}
    \label{fig:exploration.abundance}
\end{figure}
In Figure \ref{fig:exploration.abundance} we show how our key observables depend on the assumed reference abundance pattern.
In practice, the choice of reference abundance pattern has a negligible effect on our predictions. This is unsurprising, as the patterns considered are all broadly similar. By contrast, the prescription adopted for scaling nitrogen has a much stronger impact, particularly on the \bptnii\ diagram. The commonly used scalings both suppress nitrogen abundances at low metallicity (relative to a simple linear metallicity scaling) and enhance them at very high metallicity (see Figure~\ref{fig:modelling:abundance_scalings}). The net effect is to broaden the range of [N\textsc{ii}]/H$\alpha$ ratios spanned by the models.  

Directly changing the nitrogen or carbon abundances relative to oxygen primarily affects the corresponding line luminosities and ratios (not shown). The normalisation follows a sub-linear increase or decrease with abundance, with larger fluctuations at low metallicities compared to high metallicities (due to lower electron temperatures).
While other lines are less sensitive than those of nitrogen or carbon, they remain influenced by these changes. Fixing the abundance of these elements alters both the overall elemental composition and the thermal balance, leading to observable shifts in other line luminosities and ratios, particularly at high metallicities.





\subsection{Depletion and Grains}\label{sec:exploration:depletion}
\begin{figure}
    \centering
    \includegraphics[width=\columnwidth]{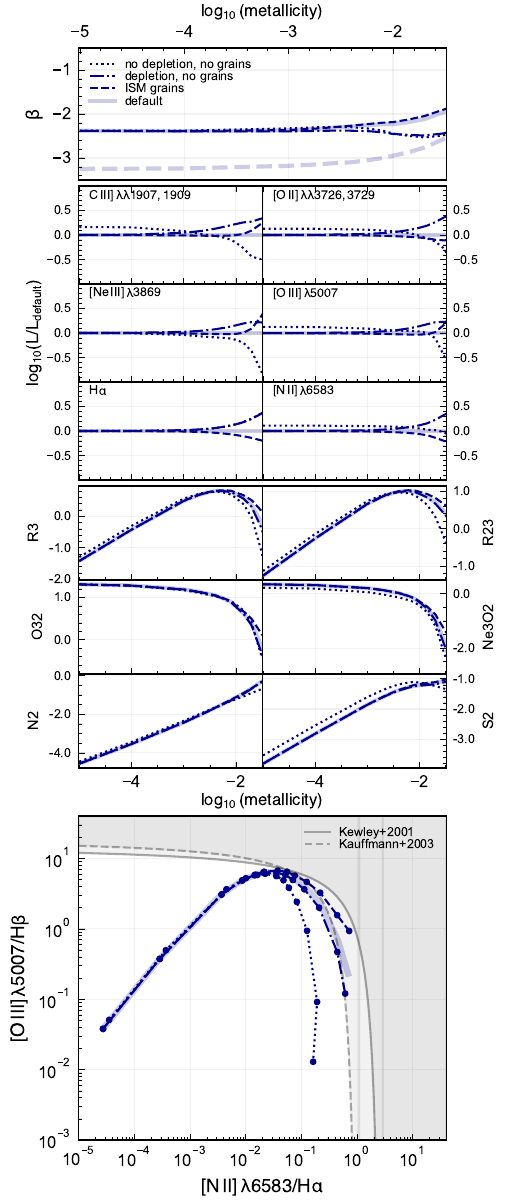}
    \caption{Same as Figure \ref{fig:exploration.sps} but showing the effect of the choice of grain mixture, including omitting grains entirely. Our default choice is Orion type grains with PAHs (see \S~\ref{sec:modelling:photoionisation:grains}). }
    \label{fig:exploration.grains}
\end{figure}
\begin{figure*}
    \centering
     \includegraphics[width=\columnwidth]{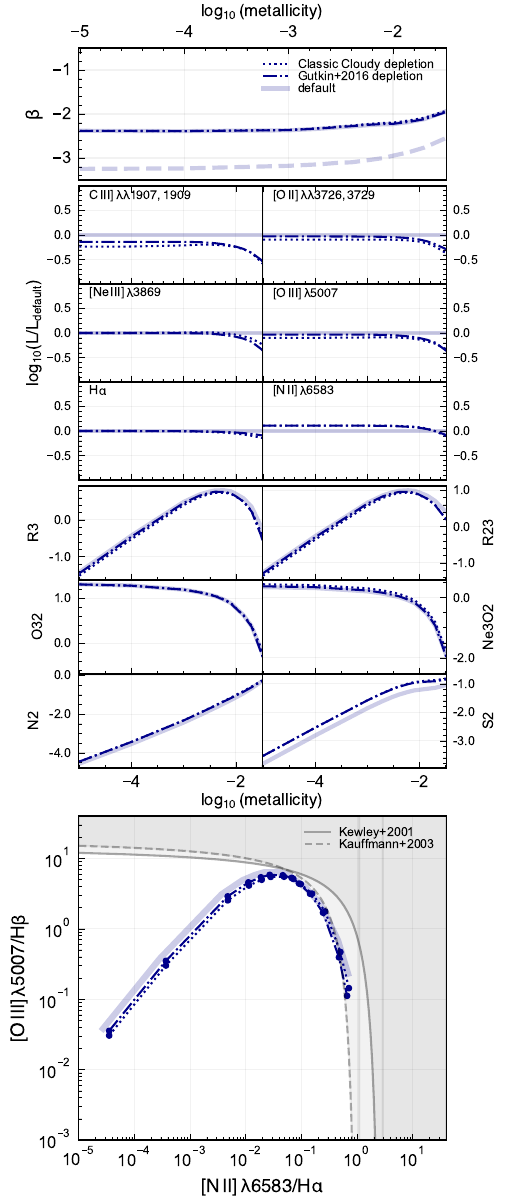}
    \includegraphics[width=\columnwidth]{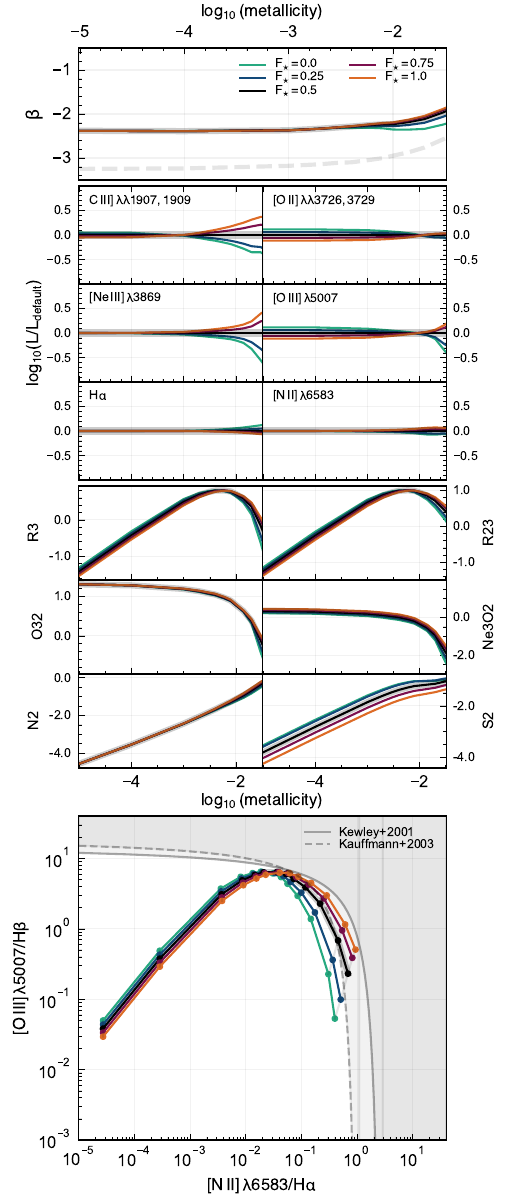}
    \caption{Same as Figure \ref{fig:exploration.sps} but showing the effect of the choice of depletion pattern (left) and depletion scale (right). We adopt the \cite{Jenkins2009} depletion formalism, setting $F_{\star}=0.5$ as our default (see \S~\ref{sec:modelling:photoionisation:depletion}).}
    \label{fig:exploration.depletion}
\end{figure*}
Finally, we explore the impact of depletion onto dust grains. Throughout this work, grains have been included by default using the \verb|Orion| mixture in \cloudy. As discussed in \S\ref{sec:modelling:photoionisation:depletion} and \S\ref{sec:modelling:photoionisation:grains}, depletion affects \hii\ regions in several ways. 
First, it removes key elements from the gas phase, thereby reducing their abundances and affecting line emission. 
Second, dust grains absorb ionising photons before they can ionise the gas, effectively reducing the ionisation parameter. 
Third, grains attenuate and redden the emergent nebular emission. 
Finally, dust grains influence the thermal balance of the \hii\ region by heating the gas through photoelectric effect and modifying cooling rates via the depletion of key elements.

To explore the impact of depletion and dust grains, we consider four scenarios in Figure \ref{fig:exploration.grains}: (i) no depletion and no grains, (ii) depletion with no grains, (iii) depletion with Orion mixture grains (our default assumption), and (iv) depletion with ISM mixture grains. We note that the scenario with depletion but no grains is not physically realistic; it is included purely to isolate the effects of elemental depletion from the additional influences of dust grains.

Compared to the cases with no grains, including dust reddens the UV continuum slope; however, this effect only becomes significant at high metallicities where grain abundance scales with carbon and silicon content. 
Hydrogen recombination lines are similarly affected: without grains, the H$\beta$ luminosity is approximately 0.2~dex higher at $Z=0.01$. Adopting the ISM grain mixture produces a modest additional reddening of the UV continuum and further attenuation of the hydrogen lines. This is because the ISM mixture contains a larger fraction of small grains, resulting in a higher dust mass absorption coefficient compared to the Orion mixture, and thus increases the attenuation for a fixed total dust mass.

The impact on metal lines, and consequently on key line luminosities and ratios, is more complex. First, depletion removes key elements from the gas, effectively reducing their abundances. Dust grains also alter the thermal balance of the \hii\ region. 
While line ratios are often constructed from pairs of nearby lines or complexes to minimise attenuation by dust, differences in both wavelength (\eg\ \hb\ is bluer than \oiii $\lambda5007$\AA) and the physical depth at which these lines originate within the \hii\ region can still lead to variations in observed ratios.
Together, these effects produce complex changes in both individual line luminosities and line ratios. One notable consequence is the shift of the high-metallicity locus in the \bptnii\ diagram toward the AGN region. This shift arises from a combination of elemental depletion and broader grain physics, particularly when the ISM grain mixture is assumed. Comparing the scenario with no grains or depletion to the ISM grain case, this effect can produce a shift of more than $0.5$~dex, emphasising the importance of accounting for dust and depletion in modelling.

We further investigate this issue in Figure~\ref{fig:exploration.depletion}, examining both the choice of depletion pattern and, in the case of the \citet{Jenkins2009} model, the effect of varying the scale parameter $F_{\star}$.  

In the left panel of Figure~\ref{fig:exploration.depletion}, we compare our key observables under alternative depletion patterns, including the ``Classic \cloudy'' prescription and the pattern adopted by \citet{Gutkin2016}. Overall, adopting these patterns leads to relatively modest changes. The most noticeable effect is a mild suppression of line luminosities at high metallicity. This primarily reflects the fact that, in these prescriptions, a larger fraction of carbon and oxygen are depleted compared to our default \citet{Jenkins2009} pattern, which affects the thermal balance (reducing the gas-phase abundance of carbon and oxygen available for cooling) in the \hii\ region and also increases extinction. 
The alternative patterns also exhibit zero depletion for nitrogen and sulphur, while our default model shows moderate depletion. 
This results in slightly elevated values for nitrogen and sulphur line strengths and ratios, which moderately affects the \bptnii\ diagram.

The right panel of Figure~\ref{fig:exploration.depletion} shows the impact of varying the \citet{Jenkins2009} depletion scale parameter $F_{\star}$. Increasing $F_{\star}$ strengthens the level of depletion, thereby amplifying the suppression of line luminosities and reddening of the UV slope, while decreasing $F_{\star}$ correspondingly weakens these effects. In other words, changes in $F_{\star}$ modulate the same trends already noted for the choice of depletion pattern, but with a smoother and more continuous scaling. 
A non-intuitive behaviour emerges at high metallicities, where line strengths actually increase as $F_{\star}$ increases (\ie\ higher depletion). 
High metallicity typically leads to lower electron temperatures due to efficient metal line cooling. However, higher metal depletion removes them from the gas phase, reducing cooling efficiency, thus raising the electron temperature and boosting these collisionally excited lines.
Another notable effect is seen in the [S\,\textsc{ii}] emission line ratio. This arises from the strong sensitivity of sulphur depletion to the choice of the $F_{\star}$ parameter. In the \citet{Jenkins2009} prescription, the fraction of sulphur locked into dust grains varies from nearly zero at $F_{\star}=0$ to $\sim 80\%$ at $F_{\star}=1$. Consequently, changes in $F_{\star}$ directly modulate the available gas-phase sulphur abundance and hence the strength of the [S\,\textsc{ii}] lines.

In summary, our parameter exploration demonstrates that while absolute emission-line luminosities are primarily dictated by the ionising photon budget (set by stellar population age, metallicity, and the IMF), diagnostic line ratios and UV continuum slopes are shaped by a complex interplay between abundance pattern, gas density, dust physics, and ionisation parameter—underscoring the critical need for self-consistent photoionisation modelling when interpreting galaxy spectra.
We also demonstrate the effect of a subset of these parameters in the following section, using galaxies from toy models and cosmological simulations.

\section{Example usage}\label{sec:examples}

We now illustrate the versatility of \synthesizer\ through six representative examples, chosen to span a wide range of astrophysical applications. These progress in complexity: beginning with the exploration of spectra stored in our default grid, followed by the construction of a simple toy galaxy model, moving on to the detailed treatment of an individual simulated galaxy, and culminating in the analysis of an entire cosmological simulation volume. In all the examples we use our default grids, unless specified otherwise. Together, these examples demonstrate how \synthesizer\ can be used both for controlled, highly idealised experiments and for large-scale statistical studies of galaxy populations.

\subsection{Grid exploration}\label{sec:examples:grid}
\begin{figure}
    \centering
    \includegraphics[width=\columnwidth]{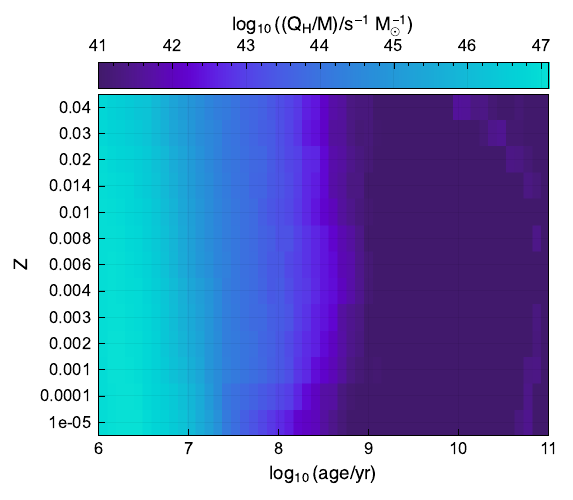}
    \includegraphics[width=\columnwidth]{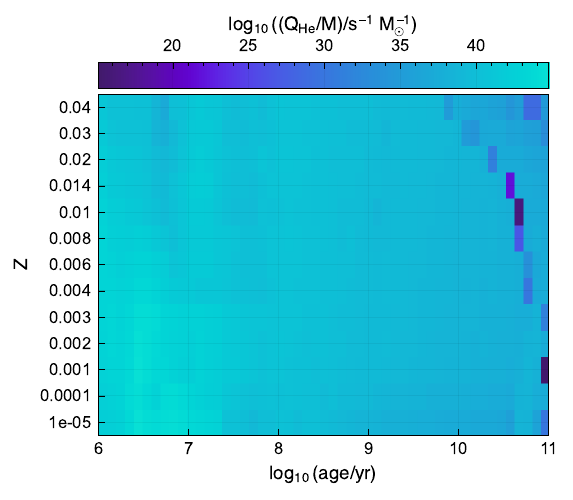}
    \caption{The distribution of the specific ionising photon production rate for hydrogen (top panel) and helium (doubly-ionised, bottom panel) for the age and metallicity range covered by version 2.2.1 of BPASS assuming a \citet{ChabrierIMF} with a high-mass cut-off of $300\, {\rm M_{\odot}}$.}
    \label{fig:examples:spectra:Q_HI_HeII}
\end{figure}

Figure~\ref{fig:examples:spectra:Q_HI_HeII} shows the ionising photon production rates for hydrogen (top panel) and doubly-ionised helium (bottom panel) across a range of ages and metallicities, based on our default SPS model and IMF \cite[][with a high-mass cut-off of $300\, {\rm M_{\odot}}$]{ChabrierIMF}.
This figure is similar to Figure~\ref{fig:modelling:specific_ionising_luminosity} (top panel), but covers the full age-metallicity range of this grid. 
It clearly illustrates that while the hydrogen ionising photon rate drops by $\approx 6$~dex over an age span of $\approx 5$~dex ($1$~Myr to $10$~Gyr), a $\approx 3$~dex ($Z=10^{-5}$ to $0.04$) variation in metallicity only results in a $\approx 1$~dex reduction. 
The helium ionising photon rate is significant at very young ages ($\sim 1$~Myr) and low metallicities ($Z \sim 0.0001$), but it drops by more than five orders of magnitude for about a dex change in age or metallicity.

\begin{figure*}
    \centering
    \includegraphics[width=\textwidth]{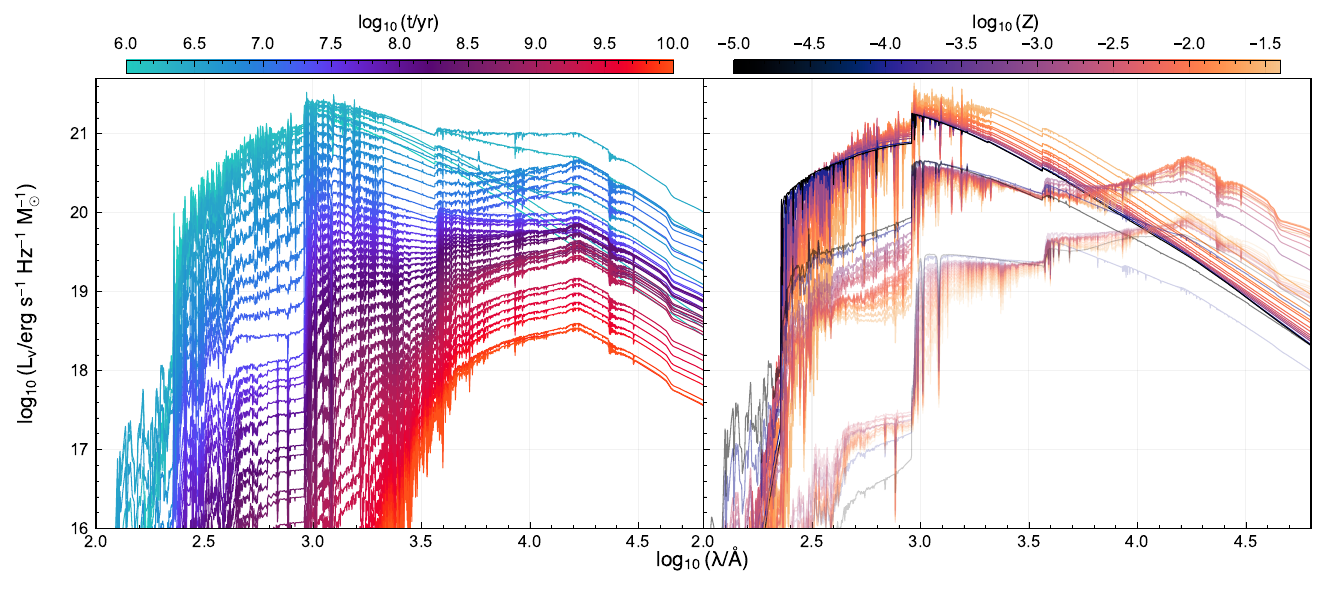}
    \caption{Pure stellar (incident) spectra in the UV to near-IR predicted by version 2.2.1 of BPASS assuming a \citet{ChabrierIMF} with a high-mass cut-off of $300\, {\rm M_{\odot}}$. The left panel shows the evolution as a function of age for $Z=0.01$. The right-hand panel instead shows the variation with metallicity for three ages, $\log_{10}(t/{\rm yr})\in\{6., 7., 8.\}$.}
    \label{fig:examples:spectra:incident_spectra}
\end{figure*}
\begin{figure*}
    \centering
    \includegraphics[width=\textwidth]{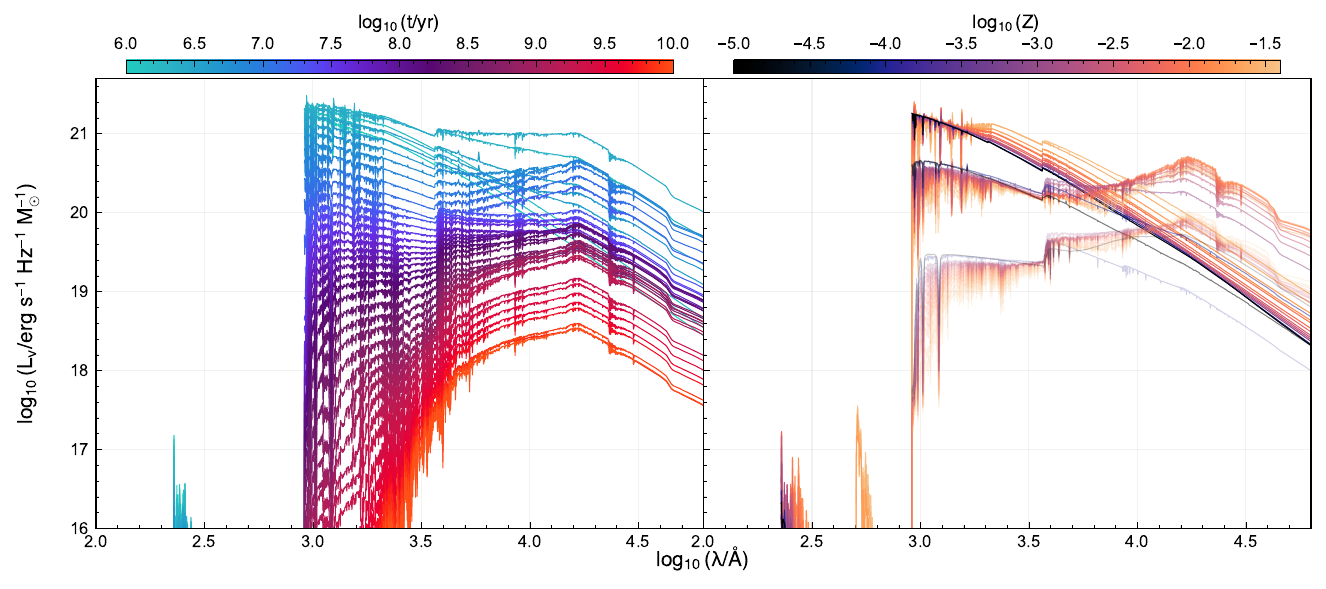}
    \caption{The same as Figure \ref{fig:examples:spectra:incident_spectra} but showing the resulting transmitted spectra for our default set of photoionisation modelling assumptions.}
    \label{fig:examples:spectra:transmitted_spectra}
\end{figure*}
\begin{figure*}
    \centering
    \includegraphics[width=0.95\textwidth]{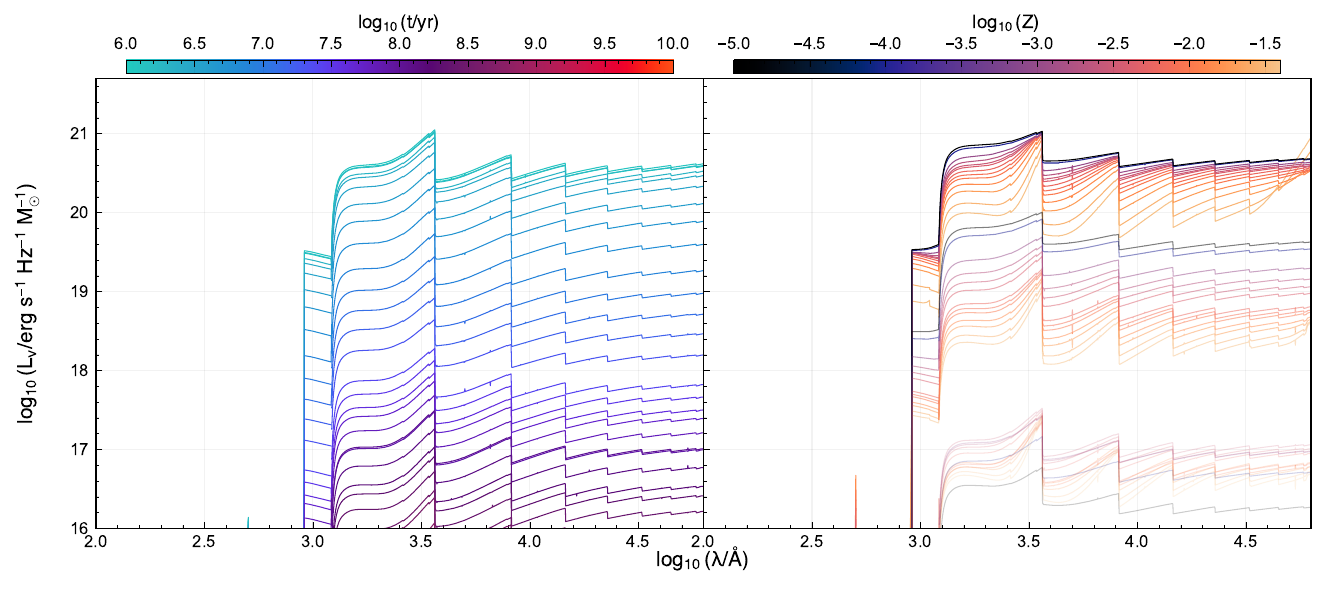}
    \caption{The same as Figure \ref{fig:examples:spectra:incident_spectra} but showing the resulting nebular continuum spectra for our default set of photoionisation modelling assumptions. }
    \label{fig:examples:spectra:nebular_continuum_spectra}
\end{figure*}

We show the incident (pure-stellar) spectra for our default grid in Figure~\ref{fig:examples:spectra:incident_spectra}. The left panel shows the spectra for an SSP as a function of age at fixed metallicity of $Z=0.01$, while the right panel shows the spectra of different metallicities for three ages, $\log_{10}(t/{\rm yr})\in\{6., 7., 8.\}$.
These spectra represent the unprocessed stellar emission before any interaction with the surrounding gas. We can see that with increasing age, the normalisation of the spectra at shorter wavelengths of the optical decreases. 

We further show the resulting transmitted spectra, which include the effects of propagation through the gas, in Figure~\ref{fig:examples:spectra:transmitted_spectra}, and the nebular continuum spectra in Figure~\ref{fig:examples:spectra:nebular_continuum_spectra}, which arise from the emission of ionised gas. 
Both sets of spectra contain little to no flux below the Lyman limit, as they have been reprocessed by the surrounding gas.
Together, these figures illustrate how the intrinsic stellar light is modified by gas and nebular processes, providing a complete picture of the SED from the incident to the processed components.

\subsection{Nebular emission from a parametric galaxy}\label{sec:examples:parametric}
\begin{figure*}
    \centering
    \includegraphics[width=0.95\textwidth]{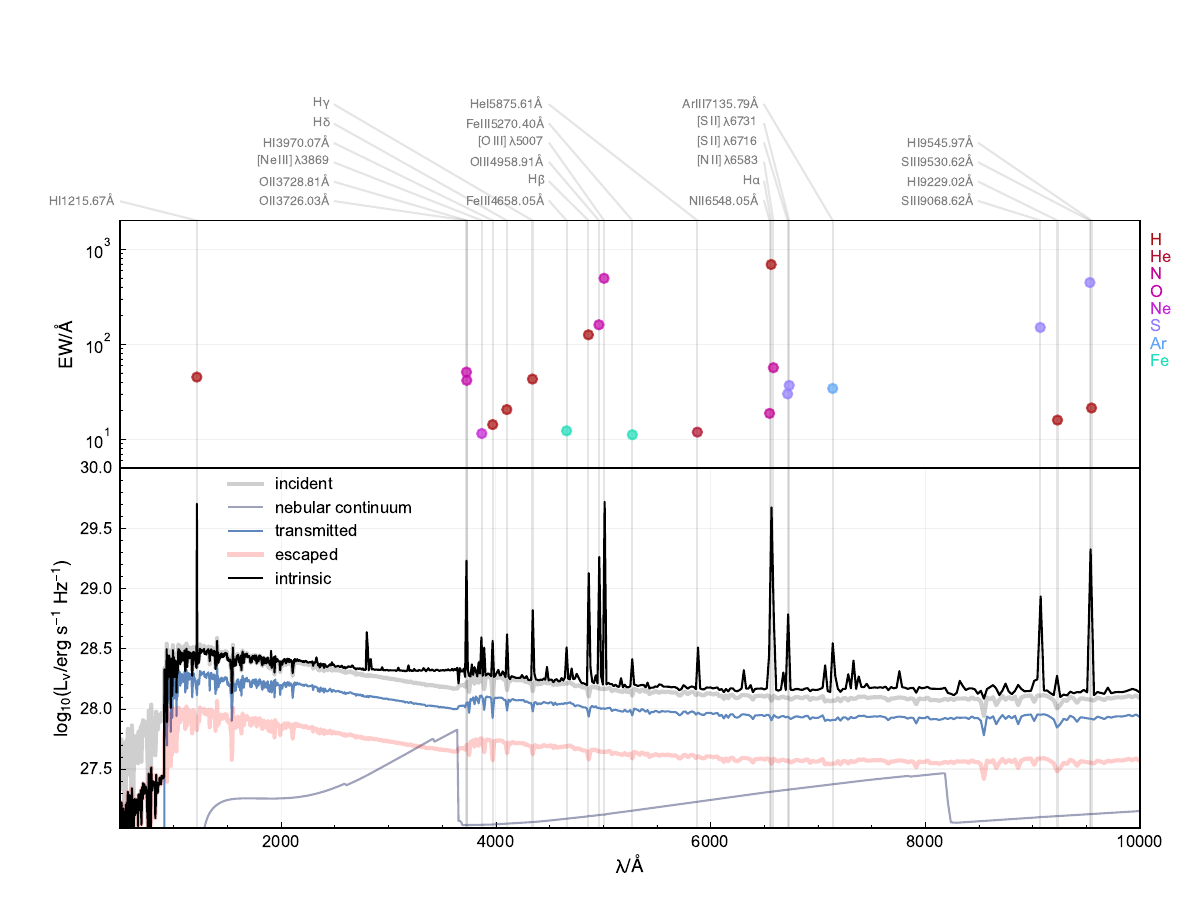}
    \caption{Spectral energy distributions and emission line equivalent widths for a galaxy formed with 50 Myr of constant star formation at $Z=0.01$. We mark emission lines with equivalent widths greater that $10$~\AA.}
    \label{fig:examples:parametric}
\end{figure*}
For our next example, we predict the SED from a toy galaxy formed under highly idealised conditions. Specifically, we assume a constant star formation history sustained over the past $50\,{\rm Myr}$ to create a total stellar mass of $10^8\, \Msun$, representative of a young, actively star-forming system, and assume a metallicity of $Z=0.01$. To account for the fact that not all ionising photons are absorbed locally, we further assume an escape fraction of $30\%$, meaning that a significant fraction of ionising radiation escapes into the surrounding medium or intergalactic space. While simple, this setup provides an instructive baseline for understanding how \synthesizer\ connects stellar population synthesis predictions to nebular line and continuum emission.

Figure \ref{fig:examples:parametric} presents two complementary diagnostics of our simple galaxy model. The bottom panel shows a set of SEDs, illustrating the different physical components that contribute to the observed spectrum. 
We show the incident pure stellar emission denoted as ``incident'', which represents the unattenuated light produced directly by stars. 
A fraction of this light escapes without interacting with the surrounding gas, labelled here as ``escaped''. 
The remainder is absorbed and reprocessed within the ionised interstellar medium, giving rise to both the ``transmitted" emission and the ``nebular continuum''. Finally, we show the total emission denoted as ``intrinsic'', which is the combination of all these contributions and corresponds to the observable spectrum of the galaxy. Note that we do not apply the effects of attenuation due to dust that can exist in the ISM of this model galaxy on the observed emission (dust within the \hii\ region is already included).
This decomposition makes clear how stellar and nebular processes jointly shape both the continuum and line emission, and highlights the importance of accounting for each component when interpreting galaxy spectra. 
The top panel shows the equivalent widths (EWs) of all emission lines stronger than $10\, {\rm \AA}$ within the displayed wavelength range, calculated for the intrinsic emission. Prominent hydrogen and oxygen lines are clearly visible alongside several fainter lines.

\subsection{Line equivalent widths}
\begin{figure}
    \centering
    \includegraphics[width=\columnwidth]{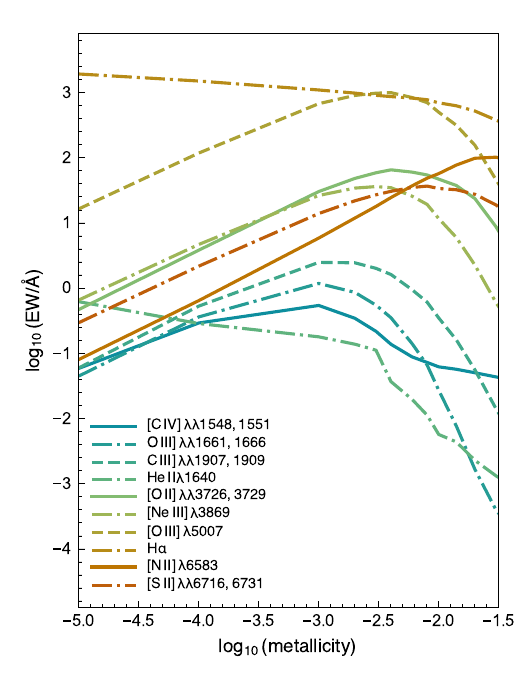}
    \caption{Equivalent width of various nebular lines as a function of metallicity generated for a parametric galaxy with a constant star formation history of 50 Myr for different metallicities.}
    \label{fig:EW_lines}
\end{figure}

Figure~\ref{fig:EW_lines} illustrates the influence of metallicity on the EWs of various lines within a parametric galaxy featuring a constant star formation history over 50 Myr. These trends mirror the behaviour of line luminosities shown in Figure~\ref{fig:exploration.metallicity}.
At metallicities lower than $Z\sim0.005$, the metal line EWs increases with metallicity due to their increased abundances; beyond this metallicity, they decrease as reduced ionising photon production rate, more efficient cooling, and increased dust attenuation become dominant. Hydrogen and helium lines show a consistent decline in EW across all metallicities because of the lower ionising photon production and higher dust attenuation associated with increasing metallicity.

\subsubsection{H$\alpha$ equivalent widths}\label{sec:examples:haew}
\begin{figure}
    \centering    
    \includegraphics[width=\columnwidth]{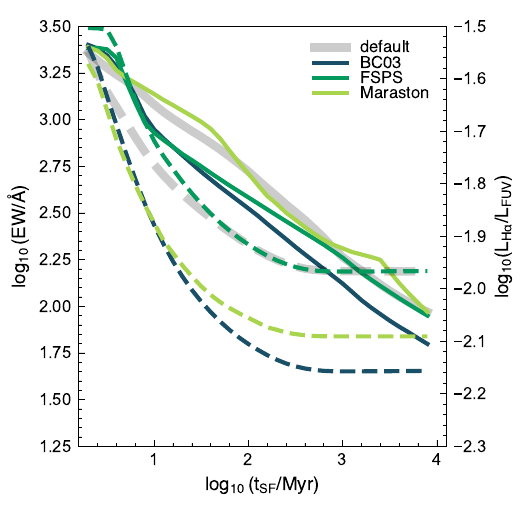}
    \includegraphics[width=\columnwidth]{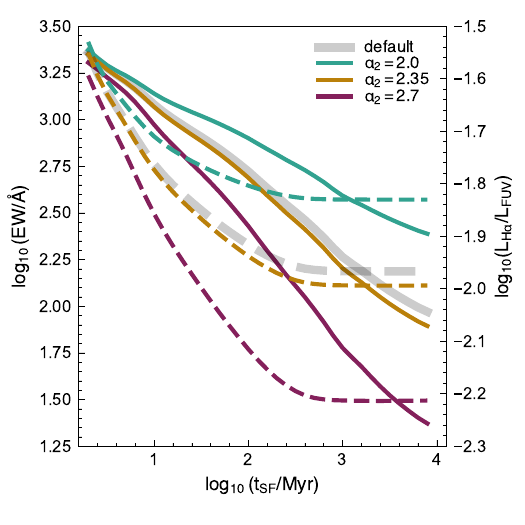}
    \caption{H$\alpha$ equivalent width (solid lines) and H$\alpha$/FUV luminosity ratio (dashed lines) as functions of star formation duration (asuming metallicity of $0.01$) for different SPS models (top panel, default model and BC03, FSPS, and Maraston) and IMF slopes (bottom panel, default model and $\alpha_{2}\in\{2.0,\,2.35,\,2.7\}$).}
    \label{fig:examples:haew_sps_imf}
\end{figure}
Observations of extremely large H$\alpha$ EWs ($>1000\,{\rm \AA}$) are now common in the high-redshift Universe, particularly in galaxies undergoing intense star formation \cite[\eg][]{Begley2026,Llerena2026}. Such values provide strong constraints on the recent star formation history and the properties of the stellar population.

Figure~\ref{fig:examples:haew_sps_imf} explores how H$\alpha$ EW depends on the duration of prior star formation across various SPS models (top panel: default model (BPASS), BC03, FSPS, and Maraston) and different high-mass slopes of the IMF (bottom panel; default model and $\alpha_{2} \in \{2.0, 2.35, 2.7\}$). These are the same set of SPS models and IMF variations explored in \S~\ref{sec:exploration:sps} and \ref{sec:exploration:imf}, respectively. 
Additionally, the same figure examines how these parameters influence the ratio of H$\alpha$ to far-UV luminosity (dashed lines), a widely used diagnostic of star formation activity and the ionising photon budget. Together, these comparisons provide insight into the physical conditions that can give rise to the extreme emission-line properties observed in galaxies at high redshift.

Figure \ref{fig:examples:haew_sps_imf} explores these dependencies for a toy galaxy with constant star formation for a duration ranging from $2\, {\rm Myr} - 10$~Gyr with metallicity $Z=0.01$ and a total stellar mass of $10^{8}\, \Msun$. 
For galaxies experiencing very young bursts of star formation, the predicted H$\alpha$ EWs are nearly identical ($\approx 2000\,{\rm \AA}$) regardless of the SPS model or IMF slope. At longer durations, however, the EWs diverge. The default model and Maraston show a similar evolution in the EW, while BC03 and FSPS decline rapidly (offset by $\approx 0.2$~dex from the default at $10$ Myr), with the decline in FSPS slowing down at longer durations to match BPASS and Maraston values. These discrepancies reflect differences in stellar template libraries, isochrones, and specific stellar populations within each model. 
Additionally, IMFs with steeper high-mass slopes decline much more rapidly, reflecting the reduced contribution of massive, ionising stars in such populations. 

Similarly, the H$\alpha$ to far-UV luminosity ratio shows distinct variations among the different SPS models and IMFs. In this case as well, the ratio declines with increasing duration, while flattening beyond $300$~Myr. Here, our default model and FSPS show similar evolution, while BC03 and Maraston exhibit a sharper decline that diverges beyond $10$~Myr. IMFs with steeper high-mass slopes decline much more rapidly.

\subsection{Impact on broadband photometry}\label{sec:examples:photometry}
\begin{figure}
    \centering
    \includegraphics[width=1\columnwidth]{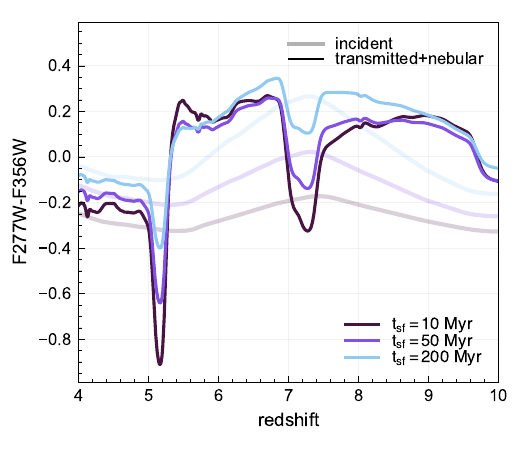}
    \caption{Predicted JWST NIRCam F277W$-$F356W colours as a function of redshift for galaxies with three different durations of constant star formation: $10$, $50$, and $200\ {\rm Myr}$.}
    \label{fig:examples:colours}
\end{figure}

In very young stellar populations, the contribution from nebular continuum and line emission can be sufficiently strong to significantly affect broad-band photometry and key spectral indices \citep[see e.g.][]{FLARES-VI}. In this example, we investigate the impact of nebular emission on the broadband photometry (colours) of high-redshift ($z>4$) galaxies.

Figure \ref{fig:examples:colours} shows the evolution of the JWST NIRCam F277W$-$F356W colour over the redshift range $z=4$--$10$ for galaxies with three different durations of constant star formation: $10$, $50$, and $200\ {\rm Myr}$, both with and without nebular emission. In the absence of nebular emission, the colours evolve smoothly, peaking at $z \approx 7.2$ when the two filters straddle the rest-frame Balmer break. Including nebular emission introduces much stronger variations in the colour evolution, driven by individual emission lines—particularly \oii, \oiii, and H$\alpha$—as they move into and out of the filter bands across redshift.

This behaviour is advantageous because it can enable not only more accurate photometric redshift estimates but also measurements of nebular line strengths, providing constraints on the star formation history, metallicity, and other physical parameters. Nevertheless, specific filter choices can introduce likelihood maxima in the photo-z estimation, potentially misidentifying galaxies at higher redshifts \cite[\eg][]{ArrabalHaro2023}.

\subsection{Individual simulated galaxy}\label{sec:examples:individual}
\begin{figure*}
    \centering
    \includegraphics[width=2\columnwidth]{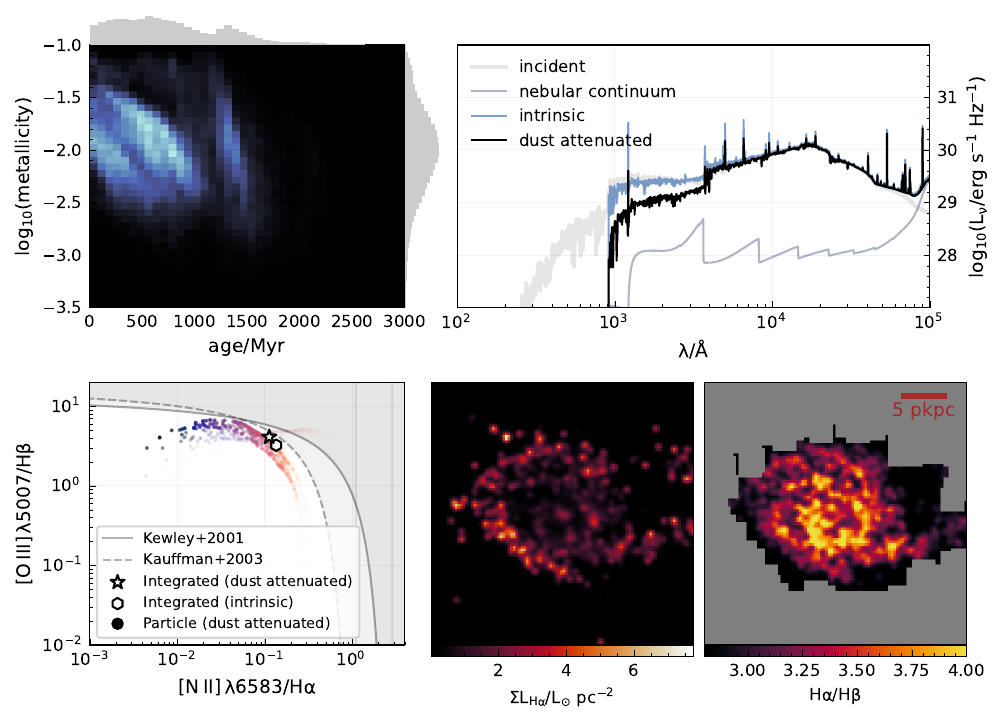}
    \caption{Clockwise from the top-left: the star formation and metal enrichment history of the galaxy; the spectral energy distributions showing the pure stellar (incident), nebular continuum, and total emission; a map of the Balmer decrement; a map of the H$\alpha$ luminosity; and the \bptnii\ diagram. In the latter, we show both the distribution of individual star particles, weighted by their observed H$\alpha$ luminosity, and the integrated galaxy position.}
    \label{fig:examples:tng}
\end{figure*}

For our next example, we apply \synthesizer\ to a single galaxy extracted from the Illustris–TNG50 cosmological simulation \cite[(50 cMpc)$^3$ volume with a gas mass resolution of $8.5 \times 10^4$~M$_{\odot}$,][]{Nelson2019, Pillepich2019}. This example demonstrates how the \synthesizer\ framework can be coupled directly to state-of-the-art hydrodynamical simulations, allowing us to move beyond idealised star-formation histories and instead model the nebular emission of galaxies with fully resolved, physically motivated assembly histories.

In the top-left panel of Figure~\ref{fig:examples:tng}, we show the binned star formation and metal enrichment history of this galaxy, revealing a complex and highly structured formation history. To incorporate the effects of dust, we apply a line-of-sight attenuation correction based on the metal surface density, following the same prescription that was used in the analyses of the \textsc{Flares} simulations \citep[see Section~2.4 in][]{FLARES-II}. As in that work, we also apply additional dust attenuation to young stars (ages less than 10~Myr) that are assumed to be embedded in their birth clouds \cite[\eg][]{CF2000}. This approach naturally introduces a variable dust optical depth for every individual star particle in the simulation, ensuring that the resulting galaxy spectrum reflects the diverse local environments within the system rather than assuming a single uniform dust screen.

For this example, we generate a suite of key observational diagnostics, including:
\begin{itemize}[leftmargin=10pt]
    \item Spectra of the galaxy showing the incident, nebular continuum, intrinsic and the dust attenuated emission. Prominent emission lines can be clearly seen in the intrinsic and dust attenuated emission, and strong dust attenuation reducing the intrinsic luminosity in the UV-optical wavelength range.
    \item Projected 2D distribution of the observed (dust attenuated) Balmer decrement (ratio of \ha\  to \hb) projected along the z-axis. We can see stronger dust attenuation around the centre of the galaxy, indicated by higher value of the Balmer decrement.
    \item Projected 2D distribution of the observed \ha\, luminosity projected along the z-axis. Following from the Balmer decrement, due to dust attenuation, there is strong reduction in the \ha\ luminosity in the centre.
    \item BPT diagram that show the distribution of individual star particles weighted by their observed (dust attenuated) \ha\, luminosity alongside the galaxy integrated location on the diagram for the dust attenuated and intrinsic SED. We can see that the position of the integrated dust attenuated and intrinsic value does not coincide, due to differential dust attenuation \cite[\eg][]{FLARES-XII}. 
\end{itemize}
Together, these outputs highlight how \synthesizer\ can connect intrinsic simulation properties directly to spatially resolved observable quantities, providing valuable insights and bridging theoretical predictions with spectroscopic surveys.

\subsection{Full cosmological simulation}\label{sec:examples:eagle}
\begin{figure*}
    \centering
    \includegraphics[width=\textwidth]{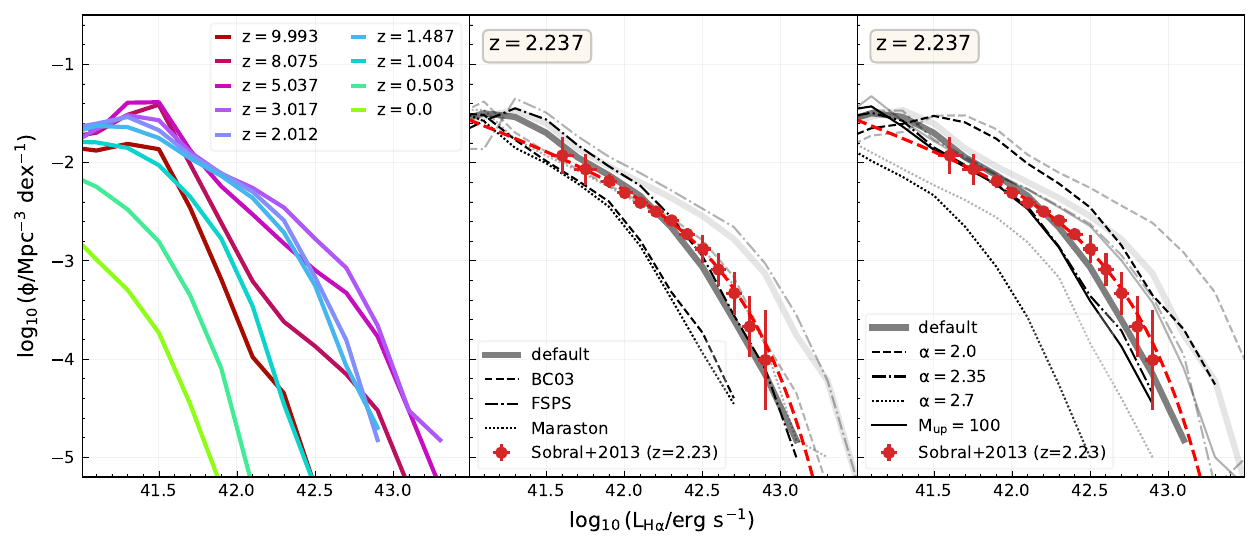}
    \includegraphics[width=\textwidth]{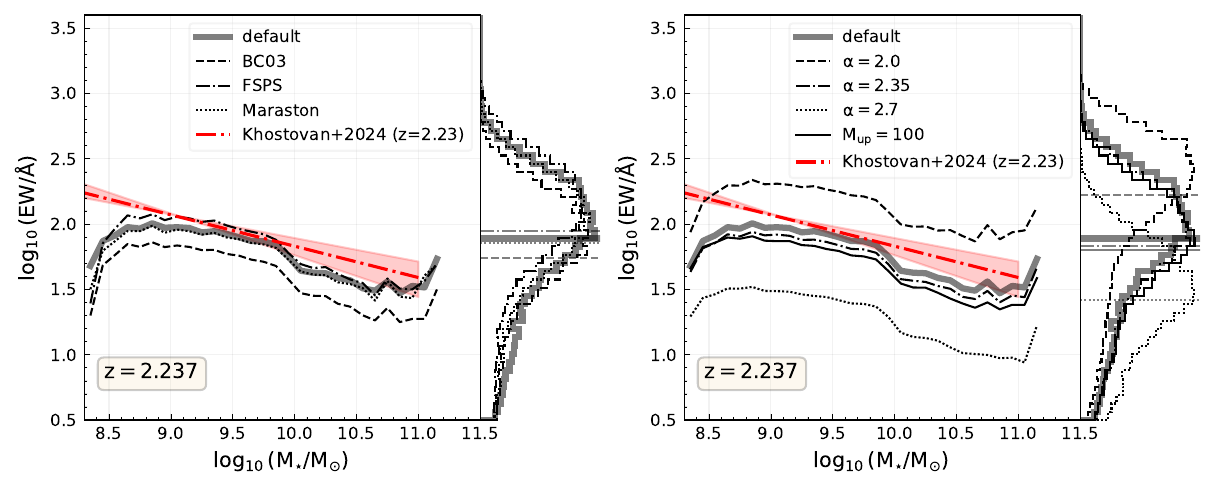}
    \caption{\emph{Top panel}: \textsc{Eagle} H$\alpha$ luminosity function for our default SPS model for $z \in [0,10]$ (left panel), for different SPS models (middle panel) and IMFs at $z=2.237$ (right panel). We compare the \ha\ LF at $z=2.237$ to observational constraints from \cite{Sobral2013} at $z=2.23$. \emph{Bottom panel}: \textsc{Eagle} \ha\,  equivalent width distribution for different SPS model (left panel) and IMFs (right panel) at $z=2.237$. We also show the normalised distribution of the EWs and denote the median value for the different models. We plot alongside the observed median relation from \cite{Khostovan2024}.}
    \label{fig:examples:eagle}
\end{figure*}
For our final example, we apply \synthesizer\ to an entire cosmological simulation---specifically, the \textsc{Eagle} cosmological hydrodynamical simulation \citep{schaye_eagle_2015, crain_eagle_2015}. We make use of the \textsc{Eagle} reference simulation, with a volume of ($100$~cMpc)$^3$ and gas mass resolution of $1.8 \times 10^6$~M$_{\odot}$. As in the TNG-50 galaxy example, we account for dust by applying an attenuation correction based on the line-of-sight metal surface density, as well as additional attenuation from birth cloud dust—consistent with the \textsc{Flares} simulation \cite[]{FLARES-I,FLARES-II}. 
Since \textsc{Flares} also utilises the \textsc{Eagle} physics model, employing this same dust attenuation model (calibrated to match the $z=5$ UVLF from \citealt{Bouwens2015} using the BPASS v2.2.1 SPS model, see Section 2.4 in \citealt{FLARES-II}) ensures a reliable recovery of observed relations for our default model.

We focus on predicting the \ha\ properties of galaxies across the full redshift range simulated by \eagle.
The top left panel of Figure~\ref{fig:examples:eagle} shows the evolution of the \ha\ luminosity function (LF) for $z\in[0,10]$ using our default SPS model. As expected \cite[\eg][]{MD2014}, the luminosity function peaks at intermediate redshift ($z \approx 2$–3) before declining sharply towards the present day.
The top-middle and top-right panels compare the predicted \ha\ LF at $z=2.237$ from various SPS models and IMFs against observational constraints at $z=2.23$ from \cite{Sobral2013}. We also show the intrinsic \ha\ LF for each model, indicated by corresponding lines of lighter shade.

As shown in Figure~\ref{fig:exploration:specific_ionising_luminosity_sps_imf}, the default model and FSPS produce a higher number of ionising photons than the BC03 and Maraston models, which is directly reflected in the intrinsic \ha\ LF. 
For the star formation and metal enrichment histories of the \eagle\ galaxies, the BC03 and Maraston models would require minimal dust attenuation to match the \cite{Sobral2013} constraints; conversely, the dust attenuated LFs for our default model (same as the SPS model used in \textsc{Flares}) and FSPS match observations closely without requiring any modification to the \textsc{Flares} prescription.
Furthermore, IMFs with a higher high-mass slope and lower high-mass cut-off underpredict the H$\alpha$ LF relative to the default model, while an IMF with a lower high-mass slope overpredicts it.

The bottom panel of Figure~\ref{fig:examples:eagle} displays the EW distribution as a function of stellar mass for \eagle\ galaxies at $z=2.237$, compared to observational data from \cite{Khostovan2024}. Unlike the \ha\ LF, the different SPS models yield only minimal differences in EW values, with medians within $\approx 0.2$~dex of the default model. The behaviour of the IMF variation mirrors that of the \ha\ LF: a lower high-mass slope results in higher EWs ($\approx 0.5$~dex offset from the default model), while a higher high-mass slope ($\approx 0.5$~dex offset from the default model) and lower high-mass cut-offs (within $\approx 0.1$~dex of the default model) result in lower EWs. 

Across this mass range, the H$\alpha$ EWs are around $100$~\AA\ for our default model, following a shallow $\propto M^{-0.3}$ decline at higher stellar masses (M$_{\star} > 10^{10}\, \Msun$). This trend is similar to that found by \citet{Khostovan2024}, though they report higher average equivalent widths, likely due to selection biases favouring high-EW sources.

\section{Summary and Outlook}\label{sec:conclusions}

In this work, we have presented the implementation of photoionised gas emission within the \synthesizer\ \cite[]{synthesizer,synthesizer-joss} synthetic observations package and illustrated its application across a range of scenarios, from simple parametric galaxies to full cosmological simulations. We have also systematically explored how key modelling assumptions—including the choice of stellar population synthesis (SPS) model, initial mass function (IMF), ionisation parameter, gas density, geometry, elemental abundance patterns, depletion, and dust physics—affect spectral diagnostics and emission-line properties within the \synthesizer\ framework. 
We summarise our findings here:
\begin{itemize}[leftmargin=10pt]
    \item SPS model: The choice of SPS model significantly affects line luminosities (Figure~\ref{fig:exploration.sps} and \ref{fig:examples:eagle}, top panels) due to variations in predicted ionising photon production rates (Figure~\ref{fig:exploration:specific_ionising_luminosity_sps_imf}, top panel). In contrast, the impact on equivalent widths and line ratios is minimal (Figure~\ref{fig:examples:haew_sps_imf} and \ref{fig:examples:eagle}, bottom panels).
    \item IMF: Varying the high-mass slope of the IMF affects both the ionising photon production rate (Figure~\ref{fig:exploration:specific_ionising_luminosity_sps_imf}, bottom panel) and the shape of the SED. This leads to significant changes in predicted line luminosities (Figure~\ref{fig:exploration.imf} and \ref{fig:examples:eagle}, top panels) and equivalent widths (Figure~\ref{fig:examples:eagle}, bottom panel). Line ratio predictions, however, remain largely unchanged by IMF variations (Figure~\ref{fig:exploration.imf}). Changing the high-mass cut-off from $300\, \Msun$ to $100\, \Msun$ for our default SPS model, produces minimal differences.
    \item Ionisation parameter ($U$): Increasing $U$ shifts the gas towards higher ionisation states (e.g., boosting $[O\textsc{iii}]/[O\textsc{ii}]$), while in the presence of dust, it increases attenuation and reddens the UV continuum ($\beta$, particularly at high metallicities), due to longer photon path lengths before reaching the ionisation front. 
    We provide two implementation modes: ``fixed" (constant across age/metallicity) and ``reference" (proportional to the cube root of the ionising photon production rate $Q_{\text{H}}$ at a reference age and metallicity). Holding $U$ fixed across age and metallicity implicitly imposes an unphysical evolution on the \hii\ region geometry due to changes in $Q_{\text{H}}$, while scaling $U$ relative to the $Q_{\text{H}}$ of a reference population dynamically tracks the stellar population's evolution (Figure~\ref{fig:modelling:ionisation_parameter}). While both approaches coincide at the reference age and metallicity, they can diverge markedly at other metallicities, driving distinct line ratio trends across parameter space (Figure~\ref{fig:exploration.ionisation_parameter}). 
    \item Gas density ($n_{\text{H}}$): Variations in hydrogen number density have a negligible impact on hydrogen recombination lines (\ha, \hb) and nebular continuum emission, but strongly affect collisionally excited metal lines with lower critical densities than $n_{\text{H}}$ as well as through more efficient cooling at higher $n_{\text{H}}$, reducing the electron temperature (Figure~\ref{fig:exploration:hydrogen_density}).  
    \item Abundance patterns and scalings: The different abundance patterns in \synthesizer\ show minimal differences in predictions (Figure~\ref{fig:exploration.abundance}). Applying specific scalings to the lines (\eg\ C/O, N/O) affects the corresponding line emission and ratios due to their altered abundances.
    \item Dust grains and depletion patterns: Dust depletion removes key species from the gas phase, altering \hii\ region abundances and thermal balance, while dust grains attenuate emergent light (Figures~\ref{fig:modelling:dust_abundance}, \ref{fig:modelling:depletion_patterns}, \ref{fig:modelling:column_density}, \ref{fig:exploration.depletion}, \ref{fig:exploration.grains}). Adopting grain mixtures with smaller grain size distributions (\eg\ ISM vs. Orion mixture, Figure~\ref{fig:exploration.grains}) or increasing depletion levels amplifies attenuation and reddening, shifting the high-metallicity BPT locus toward the AGN boundary by up to $\sim 0.5\text{ dex}$ (Figures~\ref{fig:exploration.depletion}, \ref{fig:exploration.grains}). 
\end{itemize}

Our results highlight both the sensitivity of certain observables to these assumptions and the robustness of others, providing guidance for interpreting observations and constructing realistic synthetic datasets.
As current facilities such as ALMA, \textit{Euclid}, and \jwst\ revolutionise our understanding of the Universe through high-resolution spectroscopy from the UV to the far-IR, the need for reliable and flexible synthetic tools is paramount.
Looking ahead, upcoming telescopes such as the ELT and VLT/MOONS will push these boundaries further with greater spatial and spectral resolution. In this landscape, \synthesizer\ offers a fast, flexible, and physically motivated tool to bridge the gap between theoretical models and the wealth of data from this growing era of spectroscopic missions.

\subsection{Future development and application}

\synthesizer\ is in active development, with ongoing improvements focused on expanding its capabilities and enhancing computational efficiency.

In photoionisation modelling, a key priority is enabling the use of larger and more detailed model grids. At present, the main limitation arises from the considerable storage and computational demands of these grids. To overcome this, we are developing machine-learning–based emulators that replicate the predictions of full photoionisation models with high fidelity at a fraction of the computational cost. These emulators will enable more extensive parameter exploration and facilitate rapid generation of synthetic observables.
Together with the \synthesizer\ ecosystem, they can be used for both SED fitting and simulation-based inference studies \cite[\eg][]{synference2026,Lovell2025LTU}.

Beyond these technical developments, we are currently applying \synthesizer\ to a range of state-of-the-art cosmological simulation suites such as \textsc{COLIBRE} \cite[]{Chaikin_colibre2025,schaye_colibre2025} to generate comprehensive datasets of synthetic observations. These applications will be used to systematically explore the impact of different physical and observational assumptions, and to establish a consistent framework for comparing theoretical predictions with multi-wavelength observations.

\section*{Author Contributions}

We list here the roles and contributions of the authors according to the Contributor Roles Taxonomy (CRediT)\footnote{\url{https://credit.niso.org/}}.
\textbf{Aswin P. Vijayan, Stephen M. Wilkins}: Conceptualization, Software, Data curation, Methodology, Investigation, Formal Analysis, Visualization, Writing - original draft.
\textbf{Sabrina Berger, Thomas Harvey, Christopher C. Lovell, Sophie L. Newman, William Roper, Jack C. Turner}: Software, Data curation, Methodology, Writing - review \& editing.

\section*{Acknowledgements}

We gratefully acknowledge the EAGLE and Illustris-TNG teams for their substantial efforts in developing these state-of-the-art simulations and for making their results publicly available. We have made use of the \textsc{Eagle} simulation data available on the DIRAC \href{https://cosma.readthedocs.io/}{cosma} machines.

We further acknowledge the use of several open-source software packages that enabled this work: \textsc{NumPy} \citep{numpy}, \textsc{SciPy} \citep{scipy}, \textsc{Astropy} \citep{astropy:2013, astropy:2018, astropy:2022}, \textsc{CMasher} \citep{cmasher}, and \textsc{Matplotlib} \citep{matplotlib}.

WJR, APV, and SMW acknowledge support from the Sussex Astronomy Centre STFC Consolidated Grant (ST/X001040/1). TH acknowledges support
from the ERC Advanced Investigator Grant EPOCHS
(788113).


This work used the DiRAC@Durham facility managed by the Institute for Computational Cosmology on behalf of the STFC DiRAC HPC Facility (www.dirac.ac.uk). The equipment was funded by BEIS capital funding via STFC capital grants ST/P002293/1, ST/R002371/1 and ST/S002502/1, Durham University and STFC operations grant ST/R000832/1. DiRAC is part of the National e-Infrastructure.

\section*{Data Availability Statement}
\synthesizer\ and \texttt{syncretize} are open-source, and available at \href{https://github.com/synthesizer-project/synthesizer}{https://github.com/synthesizer-project/synthesizer}\footnote{version \emph{v1.2.0}} and \href{https://github.com/synthesizer-project/syncretize}{https://github.com/synthesizer-project/syncretize}\footnote{commit \emph{8c01e75a65e2d75b7ae52923eb3337c3d36bcf18}}, respectively.

All the scripts required to generate the plots in the paper will be made publicly available on Github on the acceptance of the paper.

\bibliographystyle{mnras}
\bibliography{synthesizer-lines, flares} 



\end{document}